\documentclass[aps,twocolumn,prd,superscriptaddress,showpacs,nofootinbib]{revtex4-1}

\usepackage[utf8]{inputenc}

\usepackage{diagbox}

\usepackage{mathtools}
\usepackage{amsfonts}
\usepackage{mathrsfs}
\usepackage{bbm}
\usepackage{slashed}

\usepackage{graphicx}
\usepackage{color}
\usepackage{array}

\usepackage{blindtext}
\usepackage{placeins}
\usepackage{booktabs}
\usepackage{makecell}
\usepackage{epstopdf}
\usepackage[caption=false]{subfig}

\usepackage{xspace}
\usepackage{siunitx}
\usepackage{hyperref}
\usepackage[nameinlink]{cleveref}
\usepackage{appendix}

\usepackage{xifthen}
\usepackage{xcolor}
\hypersetup{
	colorlinks,
	linkcolor={red!75!black},
	citecolor={blue!75!black},
	urlcolor={blue!75!black}
}

\setkeys{Gin}{width=0.48\textwidth}

\graphicspath{
	{./figures/}
	{../figures/}
}

\def\pslash{p\llap{/}}
\def\qslash{q\llap{/}}
\def\kslash{k\llap{/}}

\def\s0#1#2{\mbox{\small{$ \frac{#1}{#2} $}}}
\def\0#1#2{\frac{#1}{#2}}

\newcommand{\ltpre}[1][0pt]{\mathrel{\raisebox{#1}{\scriptsize$\parallel$}}}
\newcommand{\lt}{{\ltpre[1.2pt]}}
\newcommand{\trans}{ \bot}

\def\fig#1{Figure \ref{#1}}
\def\Fig#1{Figure~\ref{#1}}

\def\Tab#1{Table~\ref{#1}}

\newcommand{\tr}{{\text{tr}}}

\newcommand{\sumint}{\int\hspace{-4.8mm}\sum}
\newcommand{\imag}{\text{i}}

\usepackage{multirow}
\usepackage{colortbl}
\definecolor{kugray5}{RGB}{224,224,224}
\newcommand{\PreserveBackslash}[1]{\let\temp=\\#1\let\\=\temp}
\newcolumntype{C}[1]{>{\PreserveBackslash\centering}p{#1}}
\newcolumntype{R}[1]{>{\PreserveBackslash\raggedleft}p{#1}}
\newcolumntype{L}[1]{>{\PreserveBackslash\raggedright}p{#1}}

\newcommand{\gettitle}{Phase structure of 2+1-flavour QCD and the magnetic equation of state}

\hypersetup{
	pdftitle={\gettitle},
	pdfauthor={Gao, Pawlowski},
	pdfkeywords={functional methods}
		{correlations functions} {phase structure}
		{functional renormalisation group} {Dyson-Schwinger equation},
	bookmarksopen=true,
	bookmarksopenlevel=2,
	bookmarksnumbered=true
}

\begin{document}

\title{\gettitle}

\author{Fei Gao}
\affiliation{Institut f{\"u}r Theoretische Physik,
	Universit{\"a}t Heidelberg, Philosophenweg 16,
	69120 Heidelberg, Germany
}

\author{Jan M. Pawlowski}
\affiliation{Institut f{\"u}r Theoretische Physik,
	Universit{\"a}t Heidelberg, Philosophenweg 16,
	69120 Heidelberg, Germany
}
\affiliation{ExtreMe Matter Institute EMMI,
	GSI, Planckstr. 1,
	64291 Darmstadt, Germany
}

\pacs{11.30.Rd, 
	12.38.Aw, 
	05.10.Cc, 
	12.38.Mh,  
	12.38.Gc 
}                             

\begin{abstract}
  We determine the chiral phase structure of $2+1$-flavour QCD in dependence of temperature and the light flavour quark mass with Dyson-Schwinger equations. Specifically, we compute the renormalised chiral condensate and its susceptibility. The latter is used to determine the (pseudo)critical temperature for general light current quark masses. In the chiral limit we obtain a critical temperature of about 141\,MeV. This result is in quantitative agreement with recent functional renormalisation group results in QCD, and is compatible with the respective lattice results. We also compute the order parameter potential of the light chiral condensate and map out the regime in the phase diagram which exhibits quasi-massless modes, and discuss the respective chiral dynamics.
\end{abstract}

\maketitle

\section{Introduction}\label{sec:Introduction}

The dynamics of QCD matter in heavy ion collisions (HIC) depends crucially on the chiral properties of QCD at finite temperature and density. For physical quark masses, QCD exhibits a smooth chiral crossover behaviour from the perturbative high temperature region of gluons and nearly massless light quarks to the hadronic phase with a sizeable spontaneous chiral symmetry breaking, for reviews see \cite{Luo:2017faz, Adamczyk:2017iwn, Andronic:2017pug, Stephanov:2007fk, Andersen:2014xxa,  Shuryak:2014zxa, Pawlowski:2014aha, Roberts:2000aa, Fischer:2018sdj, Yin:2018ejt}.

Naturally, the small current quark mass of the light up and down quarks begs the question, whether QCD under the extreme  conditions of a HIC is close to the chiral limit. A positive answer allows and supports the phenomenological access to the dynamics in HICs via systematic chiral expansions, similar to the extremely successful chiral perturbation theory ($\chi$PT) in the vacuum. The latter is based on an expansion in the pion mass that works in the regime with  quasi-massless pions. Consequently for the answer to the above phenomenologically important question we have to map out the regime with quasi-massless modes in the phase diagram of QCD.

Moreover, in the chiral limit QCD is expected to exhibit a second order phase transition, and the value $T_{c0}$ of the critical temperature, the respective universality class as well as the size of the critical regime may lead to constraints on the location of the potential critical end point (CEP) at finite density. This has led to detailed studies of the current quark mass dependence of the chiral crossover, the magnetic equation of state of QCD, for recent works with functional QCD and lattice QCD, see \cite{Ding:2019prx, Braun:2020ada, Kotov:2020hzm}. While not being fully conclusive, functional studies in low energy effective theories of QCD suggest, that the critical regime is rather small and is restricted to pion masses lower than $m_\pi\sim 1-10$ MeV, see e.g.~\cite{Braun:2007td, Braun:2009ruy, Braun:2010vd}, for a recent review \cite{Klein:2017shl}. Given, that the respective low energy effective theories encode the full chiral dynamics of QCD, it is unlikely that the inclusion of gluonic fluctuations lead to an increase of the scaling window, and this is corroborated by a recent study in functional renormalisation group (fRG) study in first principle QCD~\cite{Braun:2020ada}.

In the present work we access the magnetic equation of state within a generalised functional approach, that combines Dyson-Schwinger equations (DSE) and the functional renormalisation group (fRG). This approach was set-up and used in \cite{Gao:2020qsj, Gao:2020fbl, Gao:2021wun} for the phase structure of QCD as well as quantitative computations of QCD correlation functions in the vacuum. It extends and utilises previous computations in first principles QCD with functional approaches, for fRG works see e.g.\ \cite{Braun:2008pi, Braun:2009gm, Fister:2011uw, Mitter:2014wpa, Braun:2014ata, Rennecke:2015eba, Fu:2016tey, Rennecke:2016tkm, Cyrol:2016tym, Cyrol:2017ewj, Cyrol:2017qkl, Fu:2018qsk, Fu:2019hdw, Leonhardt:2019fua, Braun:2019aow}, for DSE works see e.g.\ \cite{Roberts:2000aa, Qin:2010nq, Fischer:2011mz, Fischer:2011pk, Fischer:2013eca, Fischer:2014ata, Eichmann:2015kfa, Gao:2015kea, Gao:2016qkh, Fischer:2018sdj, Gunkel:2019xnh, Isserstedt:2019pgx, Reinosa:2015oua, Reinosa:2016iml, Maelger:2017amh, Maelger:2018vow, Maelger:2019cbk, Aguilar:2016lbe, Aguilar:2017dco, Aguilar:2018epe}. For related lattice studies see e.g.\ \cite{Bazavov:2012vg, Borsanyi:2013hza, Borsanyi:2014ewa, Bonati:2015bha, Bellwied:2015rza, Bazavov:2017dus, Bazavov:2017tot, Bonati:2018nut, Borsanyi:2018grb, Bazavov:2018mes, Guenther:2018flo, Ding:2019prx}.

Specifically, we compute the quark and gluon propagators and the quark-gluon vertex at finite temperature. The quark propagator is then utilised to compute the renormalised light chiral condensate, whose thermal susceptibility defines the chiral crossover temperature. With these observables we discuss the magnetic equation of state in comparison to other functional and lattice results, and deduce the respective chiral transition temperature in dependence of the light current quark mass. We also compute the order parameter potential of the light chiral condensate, and discuss the regime in the phase diagram of QCD which exhibits quasi-massless modes.

In \Cref{sec:FRG-DSE} we briefly review the functional approach used here. The definition of the chiral condensate and the determination of the current quark mass parameters at the physical point are discussed in \Cref{sec:Scales}. There, we also evaluate the systematic error of the set-up. In  \Cref{sec:Results} we compute the magnetic equation of state as well as the mass dependence of the transition temperature. These results are compared to other functional QCD results as well as lattice results, also leading to an estimate for the critical temperature in the chiral limit with a combined systematic error estimate. In \Cref{sec:ChiralMassless} we derive a formula for the order parameter of the light chiral condensate, and compute it together with simple polynomial fits. Moreover, we also parameterise the current quark mass dependence of the light chiral condensate, and use these results in combination to estimate the validity regime of chiral expansion schemes. In \Cref{sec:Summary} we summarise our findings.

\section{fRG-assisted Dyson-Schwinger equations}\label{sec:FRG-DSE}

In this Section we review the functional QCD approach used here and put forward in \cite{Gao:2020qsj, Gao:2020fbl, Gao:2021wun}. In this approach the Dyson-Schwinger equations (DSEs) at finite temperature and density have been expanded about QCD within another parameter set (e.g.\ flavour, temperature, density). In \cite{Gao:2020qsj, Gao:2020fbl}, the correlation functions of two-flavour QCD in the vacuum have been used as input, obtained in the functional renormalisation group (fRG) approach, \cite{Cyrol:2017qkl}. In the present work we also utilise 2+1 flavour data from \cite{Gao:2020fbl} and the recent 2+1 flavour vacuum QCD precision data from \cite{Gao:2021wun}. The latter data are in particular used as benchmark tests.

Specifically we use 2 and 2+1 flavour vacuum data of the gluon propagator and the quark-gluon vertex. This input enables us to compute the quark propagator in the vacuum from its gap equation, see \fig{fig:QuarkDSE}. The quantitative accuracy of this result is an important self-consistency check of the approach. Then the gap equation is solved at finite temperature for different current quark masses. This requires the gluon propagator and quark-gluon vertex at finite temperature, and we expand the respective DSEs about two-flavour QCD, hence only solving for the quark mass and temperature dependence, see \fig{fig:DSEDeltaGluon} and \fig{fig:DSEDeltaQuarkGluon}.

In the following we briefly recapitulate our setup, more details can be found in \cite{Gao:2020qsj, Gao:2020fbl, Gao:2021wun}.

\subsection{Quark gap equation and chiral phase transition}\label{sec:GapEquation}

\begin{figure}[t] 
	\includegraphics[width=0.9\columnwidth]{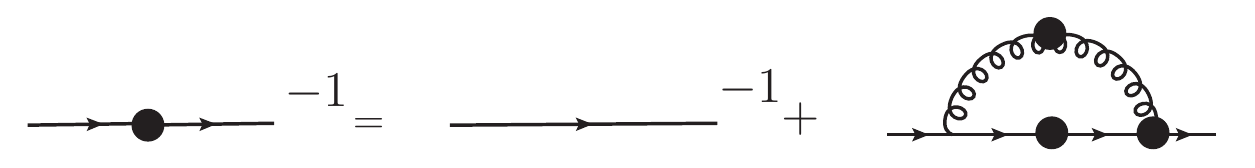}
	\caption{Quark gap equation. Lines with a blob are full propagators, that without is the classical quark propagator. The Vertex with blob is the full quark-gluon vertex, that without is the classical one. }\label{fig:QuarkDSE}
\end{figure}

The quark gap equation relates the inverse quark propagator $\Gamma^{(2)}_{q\bar q}$ to its classical counter part $S^{(2)}_{q\bar q}$, to the quark and gluon propagators, and the classical and full quark-gluon vertex, see \fig{fig:QuarkDSE}. In the vacuum we write
\begin{align}\label{eq:quark2point}
\Gamma^{(2)}_{q\bar q}(p) = Z_q(p^2)\left[\imag\,\slash{\hspace{-.193cm}p} + M_q(p)\right]\,,
\end{align}
where
\begin{align}
	\label{eq:Gn}
	\Gamma^{(n)}_{\Phi_{i_1}\cdots \Phi_{i_n}}(p_1,...,p_n)=\frac{\delta\Gamma[\Phi]}{\delta\Phi_{i_1}(p_1)\cdots \delta\Phi_{i_n}(p_n)}\,,
\end{align}
denotes the 1PI correlation functions of QCD, with $\phi=(A_\mu, c,\bar c, q,\bar q)$, for more details see e.g.\ \cite{Cyrol:2017ewj, Fu:2019hdw, Gao:2020qsj, Gao:2020fbl, Gao:2021wun}.

At finite temperature the rest frame singles out the temporal direction and we write,
\begin{align}\label{eq:quark2pointTmu}
\Gamma^{(2)}_{q\bar q}(p) =  Z^{\lt}_q(p) \Bigr[\imag\,\gamma_0 \, p_0 +  M_q(p)\Bigr]+Z_q(p)\,
\imag\,\vec \gamma \, \vec p\,,
\end{align}
where $p_0$ are now thermal Matsubara frequencies $\omega_n$,
\begin{align}
	\!\! A_\mu, c,\bar c: \ \omega_n= 2 \pi T n\,,\quad q, \bar q:\  \omega_n=&\, 2 \pi T\left(  n+\frac12\right)\,.
	\label{eq:Matsubara}
\end{align}
In \cref{eq:quark2pointTmu}, $Z^{\lt}_q$ is the wave function renormalisation for the mode parallel to the rest frame, and $Z_q=Z^{\bot}_q$, is the wave function renormalisation for the transverse modes perpendicular to the rest frame. With the parametrisation \cref{eq:quark2pointTmu}, $M_q(p)$ is defined as the pole mass. The transverse modes carry more weight in the DSE loop integrals, and we shall use the approximation $ Z^{\lt}_q/Z_q\approx 1$ in the vertices. In this approximation the difference between the longitudinal color-electric and transverse color-magnetic dressings is ignored. In \Cref{sec:SysError} the respective systematic error is discussed in detail.

Within this approximation the quark DSE at finite temperature reads,
\begin{align}\nonumber
\Gamma^{(2)}_{q\bar q}(\tilde p)-S^{(2)}_{q\bar q}(\tilde p)
=&\, \sumint\frac{dq_0}{2\pi} \! \int\frac{d^3{q}}{(2\pi)^3}\;   \Biggl[ G_{AA}{}_{\mu\nu}^{ab} (q+ p) \quad  \\[1ex]
 & \hspace{-1.5cm}\times \frac{\lambda^a}{2} {(-ig \gamma_{\mu})} G_{q\bar q}({q})
\left[\Gamma^{(3)}_{q\bar qA}\right]^b_\nu (q,- p)\Biggr]\,.
\label{eq:DSEq}
\end{align}
In \Cref{eq:DSEq}, $S^{(2)}_{q\bar q}$ is the inverse of classical quark propagator,  the full quark propagator $G_{q\bar q} = (1/\Gamma^{(2)})_{q\bar q}$ reads
\begin{align}\label{eq:QuarkProp}	
	G_{q\bar q} (q) = - \frac{Z^{\lt}_q(p) \Bigr[\imag\,\gamma_0 \, p_0 -  M_q(p)\Bigr]+Z_q(p)\,
		\imag\,\vec \gamma \, \vec p}{ \left[Z^{\lt}_q(p)\right]^2 \Bigl[p_0^2+ M_q^2(p)\Bigr]+Z^2_q(p)\,\vec p^2}\,.
\end{align}
Finally, the gluon propagator is defined by $G_{AA}= (1/\Gamma^{(2)})_{AA}$, and is detailed below, see \cref{eq:GA1} and \cref{eq:GA1}. All momenta are counted as incoming.

\Cref{eq:DSEq} can be solved with the knowledge of the gluon propagator $G_{AA}$ and the quark-gluon vertex $\Gamma_{q\bar q A}^{(3)}$. In \Cref{sec:DSEGluon} and \Cref{sec:DSEQuarkGluon} we detail how these correlation functions at finite temperature are computed on the basis of the vacuum two flavour fRG-data from \cite{Cyrol:2017ewj}.

\subsection{DSE for the gluon propagator }\label{sec:DSEGluon}

\begin{figure}[t] 
	\vspace{.3cm}
	\includegraphics[width=.95\columnwidth]{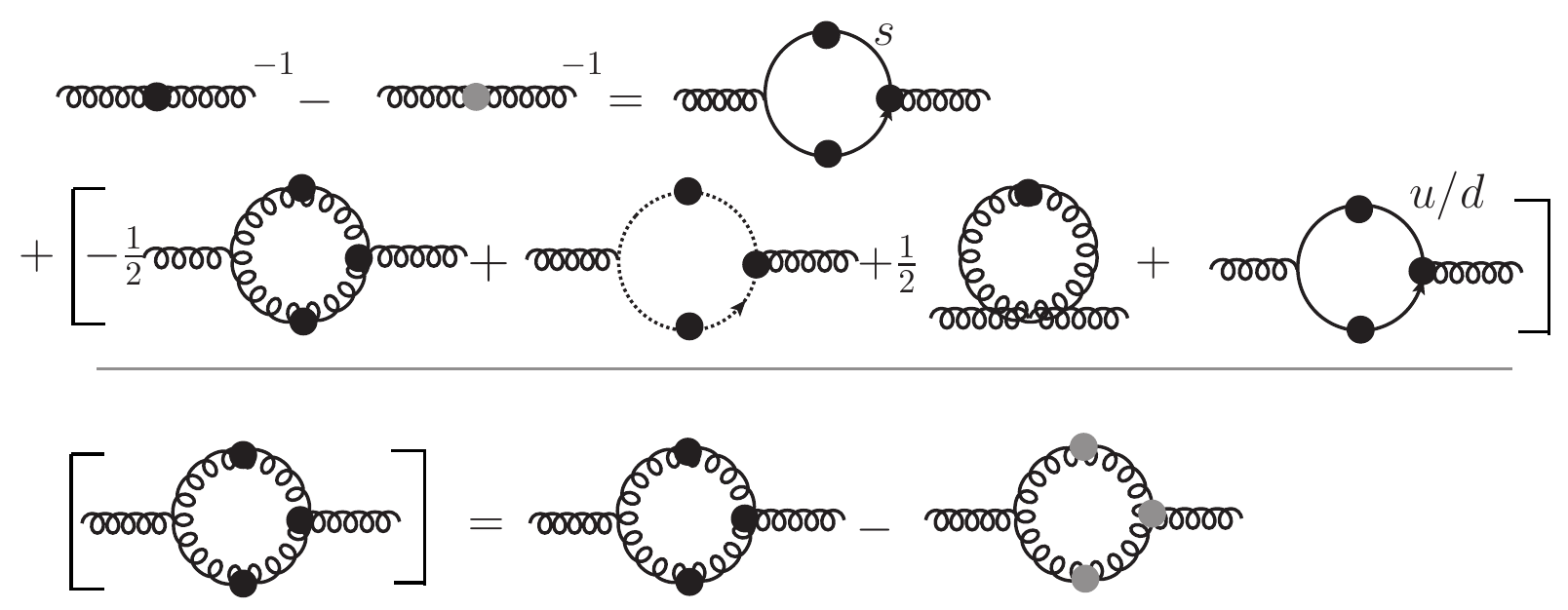}
	\vspace{-.2cm}
	\caption{Gluon DSE for the difference $\Delta\Gamma_{AA}^{(2)}$ between the full gluon propagator and the vacuum $2+1$-flavour gluon propagator. Lines and vertices with black blobs are full $N_f=2+1$ propagators and vertices at finite temperature. Lines and vertices with grey blobs are full vacuum propagators and vertices for $N_f=2$. The square bracket contains the temperature fluctuations and does not require renormalisation. We have dropped the part of the classical gluon propagator that carries the renormalisation of the strange loop. }\label{fig:DSEDeltaGluon}
\end{figure}

The DSE for the inverse gluon propagator for general flavours $N_f$ at finite temperature is expanded about that in the vacuum for two-flavour QCD. Schematically this reads,
\begin{align} \label{eq:GA}
\left. \Gamma^{(2)}_{AA}(p) \right|_{T, N_f}  =
\left. \Gamma^{(2)}_{AA}(p)\right|_{0,2} +
\Delta\Gamma^{(2)}_{AA}(p)\,,
\end{align}
and leaves us with a DSE for the difference $\Delta\Gamma^{(2)}_{AA}$ between the full inverse gluon propagator and that of vacuum $2$-flavour QCD, depicted in \Fig{fig:DSEDeltaGluon}.

At finite temperature, the gluon two-point function has color-electric and color-magnetic components,
\begin{align} \label{eq:GA1}
\Gamma^{(2)}_{AA}(p) =p^2\left[Z^M_A(p)\Pi^M_{\mu\nu}(p)+Z^E_A(p)\Pi^E_{\mu\nu}(p)
\right]+\frac{p_\mu p_\nu }{\xi}\,,
\end{align}
 with the color-magnetic dressing $Z^M_A$ and the color-electric one, $Z^E_A$. In the present work we use the Landau gauge, $\xi\to 0$. The projection operators onto the color-electric and color-magnetic directions in \cref{eq:GA1} read,
\begin{align} \label{eq:GA2}
&\Pi^M_{\mu\nu}(p)=(1-\delta_{0\mu})(1-\delta_{0\nu})\left(\delta_{\mu\nu}-\frac{p_\mu p_\nu}{\vec{p}^2}\right)\,,\notag\\[1ex]
&\Pi^E_{\mu\nu}(p)=\delta_{\mu\nu}-\frac{p_\mu p_\nu}{{p}^2}-\Pi^M_{\mu\nu}(p)\,.
\end{align}
As discussed in detail in \cite{Gao:2020qsj, Gao:2020fbl}, such difference DSEs are very stable and converge  very quickly in an iterative procedure about the initial value $\Delta\Gamma^{(n)}=0$. This can be traced back to the relatively small size of the thermal correction of the gluon propagator and quark-gluon vertex for the temperatures of interest, see e.g.~\cite{Fu:2019hdw, Braun:2020ada, Fischer:2018sdj, Gunkel:2019xnh, Gao:2020qsj, Gao:2020fbl}. The small size of the corrections also support a further approximation of the difference DSE for the gluon propagator with
\begin{align}\nonumber
&\hspace{-.6cm} T \sum_{\omega_n} \textrm{loop}_{T}(  q,p)-\int \frac{d \omega}{2 \pi}\textrm{loop}_{\textrm{vac}}(q,p) \\[1ex] \nonumber
=&
 \left[T \sum_{\omega_n}\textrm{loop}_{\textrm{vac}}(  q,p) -\int \frac{d \omega}{2 \pi}\textrm{loop}_{\textrm{vac}}(q,p)\right]\\[1ex]\nonumber
& + T \sum_{\omega_n}\Biggl[ \textrm{loop}_{T}(  q,p)-\textrm{loop}_{\textrm{vac}}(  q,p)\Biggr] \\[1ex]
\approx  &
\left[T \sum_{\omega_n}\textrm{loop}_{\textrm{vac}}(  q,p) -\int \frac{d \omega}{2 \pi}\textrm{loop}_{\textrm{vac}}(q,p)\right]\,,
\label{eq:NumStab} \end{align}
with $q_0 =\omega_n$ at $T\neq 0$ and $q_0=\omega$ at $T=0$, and $\textrm{loop}(q,p)$ stands for the loops in the second line of \fig{fig:DSEDeltaGluon}. The third line vanishes for $\Delta\Gamma^{(2)}_{AA}=0$, and is negligible for small $\Delta\Gamma^{(2)}_{AA}$. Accordingly we have dropped it, but have we have monitored its irrelevance in our explicit computation. The results for the gluon propagator in this approximation agree well with respective finite temperature results, see \cite{Gao:2020qsj, Gao:2021wun, Fu:2019hdw}, for more details we refer to these works.

Moreover, as in \cite{Gao:2020qsj, Gao:2020fbl} we drop the difference between color-magnetic and color-electric dressings: only the color-magnetic dressing $Z^E_A(p)$ is computed, and the color-electric one is approximated as $Z^E_A(p)\approx Z^M_A(p)$. This is the same approximation also used for the quark propagator, and the respective systematic error or rather is smallness is discussed in \Cref{sec:SysError}.

\subsection{Quark-gluon vertex at finite $T$}\label{sec:DSEQuarkGluon}

The DSE for the quark-gluon vertex for $2+1$-flavour QCD at finite temperature is expanded about its two- or 2+1 flavour counter part in the vacuum, computed in \cite{Cyrol:2017ewj} (2 flavour) and \cite{Fu:2019hdw, Gao:2020qsj, Gao:2020fbl, Gao:2021wun} (2+1 flavour), see also \cite{Williams:2014iea, Williams:2015cvx, Aguilar:2016lbe, Aguilar:2018epe}. We have checked, that the results of expansions about the two and 2+1 flavour vacuum results agree quantitatively. The  expansion about 2+1 flavour vacuum QCD reads,
\begin{align}
\left.\Gamma^{(3)}_{q\bar q A}(p_1,p_2)\right|_{ T, N_f}\!\! =\left.\Gamma^{(3)}_{q\bar q A}(p_1,p_2)\right|_{0, N_f}\!\! +\Delta\Gamma^{(3)}_{q\bar q A}(p_1,p_2)\,.
\label{eq:GqbarqA}\end{align}
for the light quarks, $q=l$, and the strange quark, $q=s$.

At finite temperature the vertex has color-electric and magnetic components. We identify them within our $O(4)$-symmetric approximation,  as the splitting is weak and we only keep the vacuum tensor structures. A complete basis in the vacuum with twelve basis elements is given by the eight transverse and four longitudinal projections of the Lorentz tensors
\begin{align}
\begin{array}{lcl}
\left[{\cal T}^{(1)}_{q\bar q A}\right]^\mu(p,q) =-i \gamma^{\mu}\,, &\qquad&
\left[{\cal T}^{(5)}_{q\bar q A}\right]^\mu(p,q) =i{\kslash}_+ k_-^\mu\,,\\[2ex]
\left[{\cal T}^{(2)}_{q\bar q A}\right]^\mu(p,q) =k_-^\mu\,, &\qquad&
\left[{\cal T}^{(6)}_{q\bar q A}\right]^\mu(p,q)  = i\kslash_- k_-^\mu\,,\\[2ex]
\left[{\cal T}^{(3)}_{q\bar q A}\right]^\mu(p,q) = {\kslash_-}\gamma^\mu\,, &\qquad&
\left[{\cal T}^{(7)}_{q\bar q A}\right]^\mu(p,q) =\frac{i}{2} [\pslash,\qslash]\gamma^\mu\,, \\[2ex]
\left[{\cal T}^{(4)}_{q\bar q A}\right]^\mu(p,q) =\kslash_+\gamma^\mu\,, &\qquad&
\left[{\cal T}^{(8)}_{q\bar q A}\right]^\mu(p,q) = -\frac12[\pslash,\qslash] k_-^\mu\,,
\end{array}
\label{eq:TensorsQuarkGluon}\end{align}
multiplied by the Gell-Mann matrices $\lambda^a$, see e.g.\ \cite{Cyrol:2017ewj, Gao:2020qsj}. The transverse  projection operator is denoted by $\Pi^\trans_{\mu\nu}(k)=\, \delta_{\mu\nu}- k_\mu k_\nu/k^2$, and the longitudinal one is given by $\Pi^\parallel= \mathbbm{1}-\Pi^\trans$. In the Landau gauge only the transverse part of the vertex enters the (transverse) DSEs, and hence suffices to access the full dynamics of QCD. With \cref{eq:TensorsQuarkGluon} the full transverse quark-gluon vertex is given by,
\begin{align}\label{eq:FullGqbarqA}
	\left[\Gamma^{(3)}_{q\bar q A}\!\right]^a_\mu (p, q)=\frac{\lambda^a}{2}\sum_{i=1}^8 \lambda^{(i)}_{q\bar q A}(p,q)\Pi^\bot_{\mu\nu}(k_+)\left[{\cal T}^{(i)}_{q\bar q A}\!\right]_\nu\! (p,q)\,.
\end{align}
\begin{figure}[t] 
	\includegraphics[width=0.95\columnwidth]{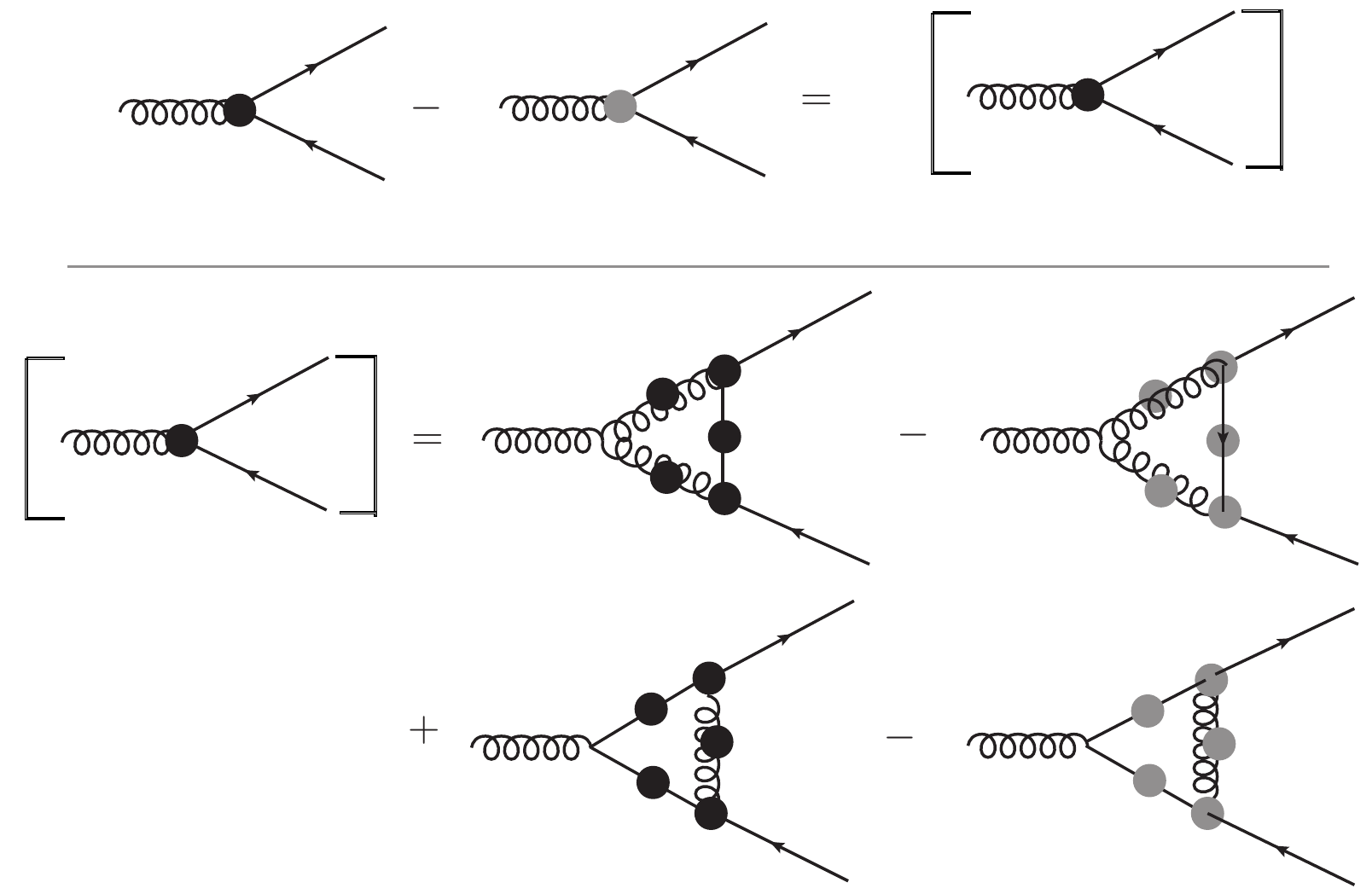}
	\caption{Quark-gluon DSE for the difference $\Delta\Gamma_{q\bar q A}^{(3)}$ between the full quark-gluon vertex (with black blob) and the vacuum $2$-flavour quark-gluon vertex (with grey blob). The temperature fluctuations in the square bracket do not require renormalisation, as in the gluon DSE \Fig{fig:DSEDeltaGluon}.}\label{fig:DSEDeltaQuarkGluon}
\end{figure}
\Cref{eq:GqbarqA} leaves us with the task of solving the Dyson-Schwinger equation for the difference. In the present work we take into account all tensor structures in the diagrams in the DSE for the quark-gluon vertex. We derive the DSE by considering the functional DSE for $\delta\Gamma/\delta A_\mu$ and then taking $q,\bar q$-derivatives. Within this hierarchy we have only consider the diagrams, that have a perturbative one-loop counterpart, and the resulting DSE is depicted in \Fig{fig:DSEDeltaQuarkGluon}. The diagrams dropped can be partly understood or taken into account as vertex dressings of the bare vertices in \Fig{fig:DSEDeltaQuarkGluon}.

In our opinion, the most prominent omission in the present approximation is the one-loop diagram with a four-quark vertex that carries (off-shell) two-quark resonances such as mesons and diquarks. Note that this omission only concerns the thermal part of the vertex, in the fRG computation underlying the input data in two-flavour vacuum QCD, \cite{Cyrol:2017ewj},  the Fierz-complete four-quark interaction vertex has been taken into account. In any case, the omission of the thermal correction of the dominant scalar-pseudoscalar channel ($\sigma$-pion) potentially leads to an underestimation of the infrared dynamics of QCD and in a linear estimate it enhances chiral symmetry breaking and hence the critical temperatures. Moreover, these contributions are relevant for the emergence of critical scaling including the values of the critical exponents in the chiral limit, see in particular  \cite{Fischer:2011pk}. Their inclusion is subject of ongoing work.

\section{Physical point and error estimates}\label{sec:Scales}

In \Cref{sec:conden} we define our order parameter for the chiral phase transition, the renormalised chiral condensate.   We use the physical point in the vacuum for setting our scales, see \Cref{sec:PhysPoint}, and discuss estimates of the systematic error of the magnetic equation of state and deduced observables such as the chiral transition temperature in \Cref{sec:SysError}.

\subsection{Chiral transition temperature and the renormalised chiral condensate}\label{sec:conden}

The chiral transition temperature is determined by the peak of the susceptibility of the renormalised light chiral condensate $\Delta_{l,R}$, with $l=u,d$ in the present isospin-symmetric approximation, defined by the thermal part of the light chiral condensate $\Delta_l$,
\begin{subequations}
	\label{eq:ChiralCond}
\begin{align}\label{eq:chiralcondren}
	\Delta_{l,R} = \frac{1}{{\cal N}_R}\Bigl[\Delta_{l}(T) -
	\Delta_{l}(0,0)\Bigr]\,.
\end{align}
with
\begin{align}\label{eq:chiralcondG}
\Delta_{l}\simeq  -\frac12  \,
T\sum_{n\in\mathbb{Z}} \int \frac{d^3 q}{(2 \pi)^3}
\tr \,G_{l\bar l} (q)\,,
\end{align}
\end{subequations}
where the factor 1/2 in \cref{eq:chiralcondG} cancels that from the sum over the light flavours, and $G_{l\bar l}$ is the light quark propagator, see \cref{eq:QuarkProp}. We remark that  the definition of the chiral condensate in \cref{eq:chiralcondG} lacks the further factor $m_l$ used in \cite{Braun:2020ada}. While such a definition ensures RG-invariance of the condensate, the present definition is that used also in chiral perturbation theory. Moreover, \cref{eq:chiralcondG} has to be renormalised, for a respective detailed discussion within the present approximation see \cite{Gao:2021wun}. For the present work the details of the renormalisation are not relevant: it drops out in \cref{eq:chiralcondren} due to the subtraction, which renders $\Delta_{l,R}$ finite. The factor ${\cal N}_R$ in \cref{eq:chiralcondren} is a convenient normalization which leaves $\Delta_{l,R}$ dimensionless, and we choose ${\cal N}_R=m^4_\pi$.

\subsection{Physical point}\label{sec:PhysPoint}

In \cite{Gao:2021wun} it has been shown, that the current approximation leads to a chiral condensate in quantitative agreement with recent lattice estimates, see e.g.\ the recent FLAG-review \cite{Aoki:2019cca}. We expect that this accuracy propagates to the renormalised chiral condensate, built on the same approximation scheme. Note however, that the difference definition eliminates the explicit dependence of $\Delta_{l,R}$ on the vacuum condensate.

For the estimate of the pion mass we apply the one-loop formula from chiral perturbation theory ($\chi$PT)~\cite{Gasser:1983yg} (valid in $2$-flavour QCD),
\begin{subequations}\label{eq:vacChiral}
\begin{align}\nonumber
	m^2_\pi=&\, M^2\left(1 -\frac{x}{2} \,\mathrm{Log}\frac{\Lambda_3^2}{M^2} + {\cal O}(x^2)\right)\,,\\[2ex]
  f_\pi=&\,f_{\pi,\chi}\left( 1 +x\,\mathrm{Log} \frac{\Lambda_4^2}{M^2}+ {\cal O}(x^2)\right)\,,
	\label{eq:fpi+mpi}
\end{align}
with the expansion parameter $x$, that depends on the mass parameter $M$. The latter is the tree-level value of the pion mass, given by the Gell-Mann--Oakes--Renner (GMOR) relation, see e.g.\ \cite{Bender:1997jf, Gao:2017gvf}, for reviews see \cite{Fischer:2006ub, Bashir:2012fs, Eichmann:2016yit}. We have,
\begin{align}\label{eq:GMORchiln}
	M^2=\,\frac{2 \, \Delta_{l,\chi} }{ f^2_{\pi,\chi}}\,m_l\,, \qquad x= \frac{M^2}{(4\pi f_{\pi,\chi})^2}\,.
\end{align}
\Cref{eq:fpi+mpi} also depends on the low energy constants
\begin{align}	\label{eq:Lambdan}
\Lambda_3  = 640(19)\,\textrm{MeV}\,,\quad \Lambda_4 = 1030(21 )\,\textrm{MeV}\,,
\end{align}
that can be computed in a Bethe-Salpeter--DSE approach. Here we simply use the recent FLAG estimates in \cite{Aoki:2019cca}, for a discussion of the dependence on the RG-scheme see \cite{Gao:2021wun}. There, the $\Lambda_n$ follow from $\bar l_n = \log {\Lambda_n}/{m_{\pi,\textrm{phys}}}$ and $\bar l_3=3.07(64)$, and $\bar l_4=4.02(45)$ for $2+1$ flavour QCD. For sufficiently light quark masses $m_l$ considered here the $ {\cal O}(m_l^3)$ is known to be small both on the lattice, \cite{Aoki:2019cca}, and in quantitative Bethe-Salpeter computations, see e.g.\ \cite{Eichmann:2016yit, Chen:2018rwz}. For the decay constant,  we utilise the FLAG results in \cite{Aoki:2019cca} for the pion decay constants in the chiral limit,
\begin{align}\label{eq:Flagfpi}
	 f_{\pi,\chi}= 86.7\,\textrm{MeV}\,,
\end{align}
\end{subequations}
in the isospin symmetric limit. The vacuum quark condensate $\Delta_{l,\chi}$ can be extracted from the UV behaviour of the corresponding constituent quark mass, see \cite{Gao:2021wun}. In the vacuum, the current approximation reduces to the the same quantitatively reliable approximation used in \cite{Gao:2021wun}. However, in the present work we expand the system of equations about the fRG vacuum solution for two-flavour QCD, while in \cite{Gao:2021wun} the DSE for the quark-gluon vertex was solved. Accordingly, the present benchmark results in the vacuum are in quantitative agreement with that in \cite{Gao:2021wun}, but not identical. Note that this remarkable agreement is yet another self-consistency check of functional methods. In the chiral limit we find,
\begin{align}\label{eq:ChiralCondchi}
 \Delta_{l,\chi}(\mu) =\, (308.9(8)\,\textrm{MeV})^3 \,,
\end{align}
in quantitative agreement with the lattice estimate $\Delta_{l,\chi}(\mu) =\, 315(6)\,\textrm{MeV})^3$ and the DSE results in \cite{Gao:2021wun}. The above relations allow us to fix the physical point: the current quark masses $m_q(\mu)$ with an RG scale of $\mu=40$\, GeV are fixed such that we have in the isospin symmetric limit with $m_l=m_u=m_d$,
\begin{subequations}
	\label{eq:physpoint}
\begin{align}\label{eq:physpointOb}
	m_{\pi,\textrm{phys}} = 138\,\textrm{MeV}\,,\qquad \frac{m_{s,\textrm{phys}}}{m_{l,\textrm{phys}}}=27\,.
\end{align}
This leads us to
\begin{align}\label{eq:physpointQuark}
	m_{l,\textrm{phys}} (\mu) = 2.49 \,\textrm{MeV}\,, \qquad m_{s,\textrm{phys}} (\mu) = 66.7\,\textrm{MeV}\,,
\end{align}
with the prediction
\begin{align}\label{eq:fpiPhys}
 f_{\pi,\textrm{phys}} =92.4\,\textrm{MeV}\,,
\end{align}
\end{subequations}
in line with the physical value and the lattice estimate in \cite{Aoki:2019cca}. We emphasise that this perfect agreement originates in the quantitative agreement of the chiral condensate in the chiral limit, $\Delta_{l,\chi}(\mu)$ in \cref{eq:ChiralCondchi}, from the present DSE computation with the respective lattice estimate. As discussed in \cite{Gao:2021wun}, $\Delta_{l,\chi}(\mu)$ is the benchmark observable in the vacuum which shows the quantitative reliability of the present approximation of the DSE and fRG in \cite{Cyrol:2017ewj}. Additionally, the correct value for $f_{\pi,\textrm{phys}}$ is a test of the 2-flavour approximations used in \cref{eq:fpi+mpi}.

\subsection{Systematic error estimate}\label{sec:SysError}

The present functional approach, or rather the approximation level of the systematic vertex expansion used here, has passed successfully numerous benchmark comparisons,  a systematic comparison can be found in \cite{Gao:2021wun}, a discussion of the systematic error estimate is done in \Cref{sec:SysError}. Apart from the impressive quantitative agreement of benchmark observables such as the chiral condensate, results of different functional approaches that constitute different resummation schemes also agree quantitatively. In conclusion we consider the approximation level used here as sufficiently converged in the vacuum.

In turn, at finite temperature we use additional approximations. In the following we discuss the ensuing systematic error of the present approximation level. The most relevant sources for the systematic error are,
\begin{itemize}
\item[(i)] No feedback of the order parameter potential: feeding back the order parameter potential into the diagrams encodes the critical dynamics beyond mean field, and we expect deviations from the full result in the scaling regime for $m_l\to 0$.
\item[(ii)] No thermal splits in the vertices, most prominently in the quark-gluon vertex: we identify electric and magnetic dressings and compute this uniform dressing by magnetic projections.
\end{itemize}
Let us first discuss the impact of \textit{(i)}: The quantitative access to the critical dynamics for $m_l\to 0$, and hence $m_\pi\to 0$, requires the inclusion of a dynamical order parameter potential. Heuristically speaking the latter takes into account the multi-quark scatterings of the resonant pion channel that become relevant in the limit of infinite correlation length. This approximation is well-tested in the vacuum, and the neglection of the chiral dynamics encoded in the higher multi-quark scattering vertices is negligible there. In the quantitative fRG-studies \cite{Mitter:2014wpa, Cyrol:2017ewj}, the higher-order scatterings have been taken into account systematically via dynamical hadronisation in terms of the scattering of the respective resonant interaction channels, for a recent review see \cite{Dupuis:2020fhh}. This allows a quantitative access to the critical chiral dynamics, and the effects on vertices and propagators are very small: The respective changes are well within the systematic error estimates for the given approximations. Moreover, the results for quark and gluon propagators as well as the quark-gluon vertex with the full order parameter potential are in quantitative agreement with that of the DSE-approach used here, see \cite{Gao:2021wun}. In the context of the present work this is most impressively confirmed by the chiral condensate in the chiral limit, $\Delta_{l,\chi}$ in \cref{eq:ChiralCondchi}, which is in quantitative agreement with other benchmark computations. In conclusion, the critical dynamics does not affect the results in the vacuum, and the scaling regime for $m_l\to 0$ is very small.

At finite temperature these findings strongly suggest that the higher order multi-quark scatterings only play a r$\hat{\textrm{o}}$le in the critical scaling regime for temperatures $T$ close to the chiral phase transition temperature  $T_{c0}$ in the chiral limit and small light current quark masses $m_l\to 0$. These higher scattering processes have been taken into account in \cite{Braun:2020ada} for pion masses $m_\pi\geq 30$\,MeV without any sign of criticality. Taking also into account the small difference between mean field scaling and full $O(4)$ or $O(2)$ scaling, we estimate that the systematic error of neglecting the higher order multi-quark scatterings is negligible also for $m_\pi<30$\,MeV. Note that this estimate applies to the absolute value of observables such as $T_c$ as a function of the light quark mass. As already emphasised at the end of \Cref{sec:DSEQuarkGluon}, the approximation \textit{(i)} has to be improved for the access to critical scaling, and in particular for a computation of the critical exponents, see \cite{Fischer:2011pk}. This is subject of ongoing work.

In summary, we are led to a systematic error estimate for \textit{(i)}, which is subleading for the observables studied here, and is dominated by other approximation effects.

We proceed with the evaluation of \textit{(ii)}, the missing thermal splits in propagators and vertices. This affects the contribution of the zeroth Matsubara frequency $\omega_0$, that is either $\omega_0=0$ for the gluons and ghost, and $\omega_0 = \pi T$ for the quarks. There, the spatial momentum dependence is changed for $\vec q{\,}^2 \lesssim (2 \pi T)^2$. In turn, the thermal effects for higher Matsubara frequencies are negligible. Within diagrams, this thermal change of correlation functions is suppressed with the measure factor $\vec q{\,}^2$ of the spatial momentum integration. This combination of the decay of thermal contributions for $\vec q{\,}^2\to\infty$ and its phase space suppression for $\vec q{\,}^2\to 0$, suggests the feasibility of $O(4)$-symmetric approximations. Here, $O(4)$ refers to the Euclidean space-time symmetry in the vacuum. In the present work we implement this idea with using projections on the magnetic (spatial) dressings of propagators and vertices. Then, the electric dressing is identified with the magnetic one. The above line of reasoning has been checked within different theories within the functional renormalisation group approach, \cite{Dupuis:2020fhh}, including Yang-Mills theory at finite temperature, \cite{Cyrol:2017qkl}. Finally, we have extended the approximation used in the present work with the thermal splits for vertices and propagators, which will be discussed in detail in a forthcoming work. The respective results corroborate the arguments above and lead to negligible modifications of the observables including the chiral condensate. Moreover, the crossover temperature (for physical quark masses) only increases by less than 3 MeV, \cite{Gao:2021}.

Finally, we discuss the systematic error estimate for such an approximation. To begin with, the above suppression of thermal contributions for $\vec q{\,}^2\lesssim (2 \pi T)^2$ is undone for correlation functions, whose functional relations(fRG or DSEs) have sizeable infrared contributions for $\vec q{\,}^2/m^2_\textrm{gap}\to 0$. Here,  $m^2_\textrm{gap}$ is the characteristic mass scale of QCD: below this mass scale, momentum fluctuations do not contribute significantly to the physics. This entails that $m^2_\textrm{gap}$ is triggered by the gluon mass gap as well as he constituent quark mass. Note however, that such a survey also has to take into account effective degrees of freedom, and in particular the pion. It is here, where \textit{(i)} comes into play. Again, from detailed studies in the vacuum we know that the characteristic scale for fluctuations for the quark mass function and consequently chiral condensate is $m_\textrm{gap}\approx 1$\, GeV. This structural analysis is well confirmed within fRG-studies, where momentum fluctuations are resolved iteratively momentum-shell by momentum-shell for $\vec q{\,}^2 \approx k^2$, the latter being the infrared cutoff scale: the infrared decay of momentum fluctuations is clearly visible. The above argument also entails, that a phenomenological infrared enhancement of vertices, as often used in functional studies, has to be taken with a grain of salt as it may over-enhance thermal (and chemical potential) effects.

Apart from these structural arguments the present setup offers several simple self-consistency checks: Firstly, instead of using the magnetic projection for the computation of the $O(4)$-symmetric dressing we can either average over all Euclidean directions or we even take the electric projection. We have checked that this only leads to variations of the chiral transition  temperature of a few MeV, indeed safely below $5$\,MeV. Secondly, the slope or curvature of the chiral transition temperature  $T_c(\mu_B)$ at vanishing baryon chemical potential, $\mu_B=0$, is a direct check of the correct  infrared dynamics at finite temperature and density: despite the differences, i.e. the lack of direct density contributions to gluons and further correlations with vanishing baryon number, thermal and density effects share the same tendencies. In particular, increasing the infrared vertex strength of the quark-gluon vertex increases the size of thermal and chemical potential effects. Within the present approximation level of the DSE approach to QCD that underlies the current study, it is quantitatively agreeing with the lattice prediction, \cite{Gao:2020fbl, Gao:2020qsj}. This also holds true for the related fRG study in \cite{Fu:2019hdw}, which underlies the investigation of the magnetic equation of state in \cite{Braun:2020ada}.

In summary, the Euclidean $O(4)$-symmetric approximation used here is sound, and the systematic error estimate is achieved by applying an artificial infrared enhancement to the quark-gluon vertex that leads to a variation of the chiral transition temperature $T_c$ at physical quark masses of $5$\,MeV. The respective systematic error band is then shown in our depiction of the results for the chiral transition temperature, see \Cref{fig:Tc} and \Cref{fig:TcMpi}.

\section{Magnetic equation of state}\label{sec:Results}

A first main result is given by the chiral transition temperature $T_c(m_l)$ in dependence of the ratio $H=m_l/m_s$ of current quark masses, that can be derived from the magnetic equation of state: the transition temperature is obtained from the peak of the susceptibility of the renormalised light chiral condensate defined in \Cref{sec:conden}. The results are discussed in \Cref{sec:MagEoS}, including a comparison with the respective lattice results, \cite{Ding:2019prx, Kaczmarek:2020err, Kaczmarek:2020sif, Karsch:2019mbv, Kotov:2021rah}, as well as the functional QCD study (fRG) in \cite{Braun:2020ada}. For the comparison with the latter results we also present $T_c(m_\pi)$. This requires $m_\pi(m_l)$, for which we use \cref{eq:fpi+mpi}, valid at one-loop up to $O(x^3)$. The latter higher order terms are known to be small for the pion masses $m_\pi\leq  140$\,MeV considered in \cite{Braun:2020ada}. In \Cref{sec:Constraints} we discuss  constraints on the equation of states as well as the merits of future combined studies, utilising both, functional results as well as lattice simulations.

\subsection{Transition temperature and transition line}\label{sec:MagEoS}

The chiral transition temperature is computed from the peak of the renormalised light quark susceptibility $\chi^{(l,R)}_M$, given by,
\begin{align}
	\label{eq:MagSuscept}
	\chi_M^{\ }= \chi^{(l,R)}_M = -\frac{\partial \Delta_{l,R} }{\partial m_l} \,.
\end{align}
With \cref{eq:MagSuscept} we define the chiral transition temperature as the peak position of $\chi^{(l,R)}_M$,
\begin{align}\label{eq:Tc}
	T_c= T^{(l,R)}_c= T_\textrm{peak} \,,\quad  \quad
	\left. \frac{\partial \chi^{(l,R)}_M}{\partial T}\right|_{T=T_\textrm{peak}}=0\,.
\end{align}
In \cite{Braun:2020ada}, also the magnetic susceptibilities for the light chiral condensate $\chi^{(l)}_M$ and the reduced chiral condensate $\chi^{(l,s)}_M$ have been considered. There, it has been shown, that the respective transition temperatures $T_c^{(i)}$ with $(i)=(l)\,,\, (l,R)\,,\, (l,s)$ agree well in the crossover regime. Evidently, $T_c^{(l)}=T_c^{(l,R)}$. The second approximate relation, $T_c^{(l,s)} \approx T_c^{(l,R)}$ follows from the fact, that the transition temperatures only vary within the width of the crossover in the first place, and the thermal dependence of the strange quark condensate is far reduced. Moreover, all transition temperatures agree trivially in the second order regime.

\begin{figure}[t] 
	\includegraphics[width=.98\columnwidth]{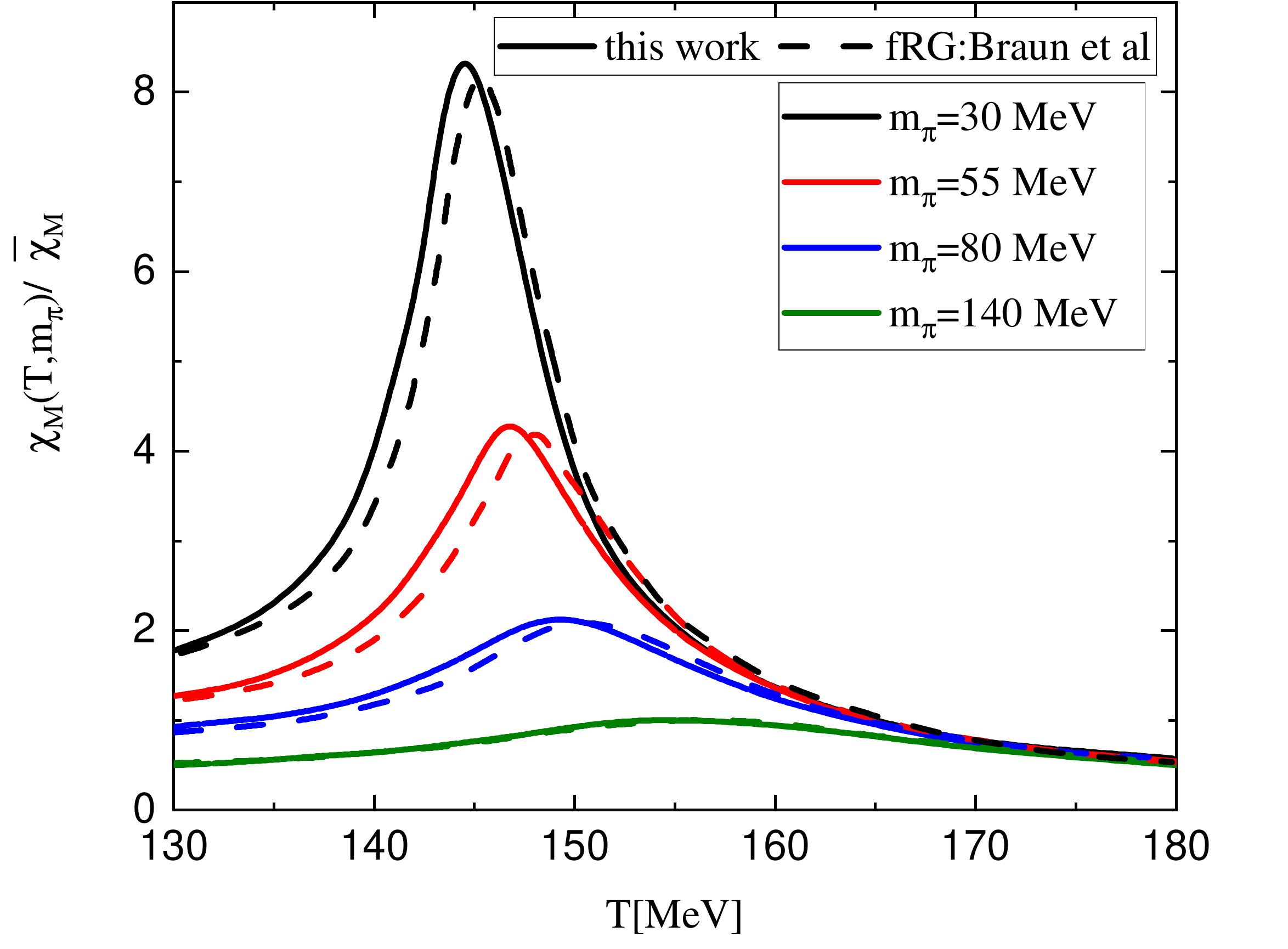}
	\caption{Temperature dependence of the renormalised magnetic susceptibility $\chi^{(l,R)}_M$, \cref{eq:MagSuscept}, for various pion masses. The results are compared to the respective fRG-results \cite{Braun:2020ada}.}\label{fig:sus}
\end{figure}
We have benchmarked our scales in the vacuum with the physical point with a isospin-symmetric pion mass,
$m_{\pi,\textrm{phys}}=138$\,MeV, see \cref{eq:physpoint}. The chiral transition temperature at the physical point now follows as
\begin{align}\label{eq:TcPhys+error}
	T_{c}=154.7^{+5}_{-5}\, \textrm{MeV}\,,
\end{align}
with the systematic error estimate \textit{(ii)} discussed in \Cref{sec:SysError}. It encodes the systematic error of the strength of the quark-gluon coupling. The latter has been varied such that $T_c( m_{\pi,\textrm{phys}})$ for physical pion masses with $\pm$5\,MeV.

\begin{figure}[t] 
	\includegraphics[width=.92\columnwidth]{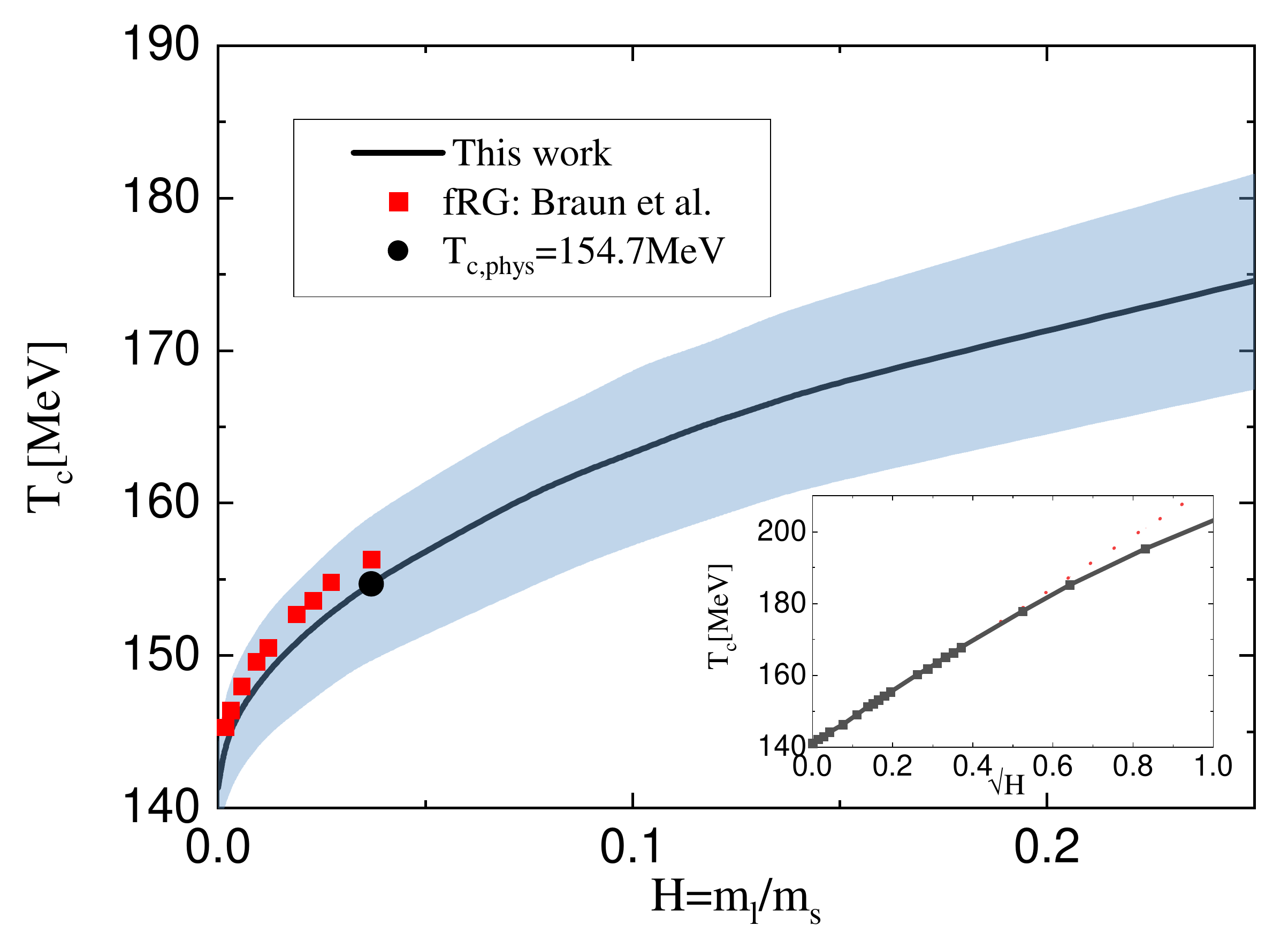}
	\caption{Chiral transition temperature for $N_f=2+1$ flavours as a function of $H=m_l/m_s$ in comparison to the fRG-results from  \cite{Braun:2020ada}. The crossover temperature $T_c$ is defined by \cref{eq:Tc}. The inset shows $T_c$ as a function of $\sqrt{H}$. Both curves are well approximated by $\sqrt{H}$ for $H\lesssim 0.25$. The blue error band is a systematic error estimate on the DSE results: it is obtained by varying the overall strength of the quark-gluon coupling such that  $T_c(m_\pi)$ changes in a range of  $T_c=155\pm5$ MeV.}\label{fig:Tc}
\end{figure}

In \fig{fig:sus} we depict the chiral susceptibilities for $m_\pi=30,55,80,140$\,MeV   in comparison to the fRG results in \cite{Braun:2020ada}. The respective  transition temperatures are depicted in \Fig{fig:Tc} as a function of $H=m_l/m_s$. One clearly sees the $\sqrt{H}$-behaviour. Note that different definitions of the transition temperature lead to minor changes of the location of phase transition in the crossover regime, but they all agree in the second order regime in the chiral limit. We also depict the pion mass, obtained in the approximation \cref{eq:fpi+mpi}, as a  function of current quark mass in \Fig{fig:Tc}.

\begin{figure}[t]
	\includegraphics[width=.99\columnwidth]{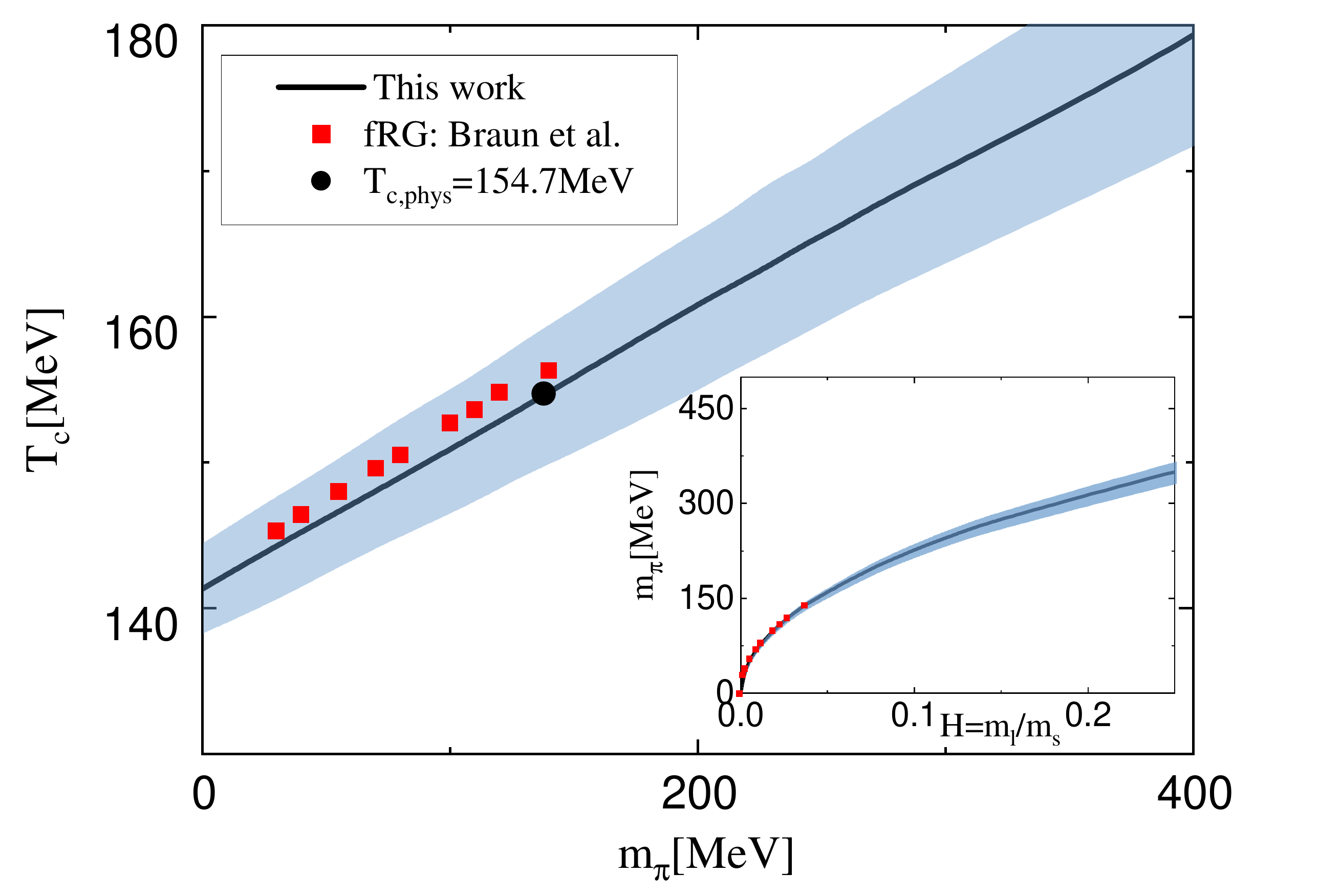}
	\caption{Chiral transition temperature for  $N_f=2+1$ flavours  as a function of the pion mass in comparison to the fRG-results from  \cite{Braun:2020ada} . $T_c\propto m_\pi$ is a very good approximation  for $H\lesssim 0.25$. The inset shows the pion mass as a function of $H=m_l/m_s$ within $\chi$PT, see \cref{eq:vacChiral}. The blue error band is a systematic error estimate on the DSE results: it is obtained by varying the overall strength of the quark-gluon coupling such that  $T_c(m_\pi)$ changes in a range of  $T_c=155\pm5$ MeV. }\label{fig:TcMpi}
\end{figure}

In the present approximation to the $m_l$-dependence of the pion mass, \cref{eq:fpi+mpi}, evidently $m_\pi\propto  \sqrt{H}$ holds close to the chiral limit. In turn, a linear dependence $m_\pi\propto H$ emerges for asymptotically large light current quark masses. Together with the approximate, but quantitative, $\sqrt{H}$-behaviour of the chiral transition temperature, this leads us to the linear dependence $T_c\propto m_\pi$ close to the chiral limit. The linear dependence is clearly visible in \Fig{fig:TcMpi}, and holds true even up to $H=1$, that is $m_l=m_s$.

In the scaling regime of QCD close to the chiral limit, the chiral phase transition temperature should show critical scaling with $T_c\sim H^{1/(\beta\delta)}$. Here, the critical exponents $\beta,\delta$ determine the universality class of QCD in the chiral limit: $m_s$ fixed, $H\to 0$.

A detailed discussion in view of the universality class of QCD in the chiral limit of the functional QCD results in the present work and in  \cite{Braun:2020ada}, as well as the lattice results of \cite{Ding:2019prx, Kotov:2021rah} is given in \Cref{sec:ChiralMassless}. For the time being, we just remark that the approximation to functional QCD with the fRG in \cite{Braun:2020ada} incorporates criticality in terms of the dynamics of the order parameter and the order parameter potential. This has been tested both in generic low energy effective theories and in QCD, see \cite{Dupuis:2020fhh}. Consequently, if critical scaling is present in QCD, it is well-captured within the approximation to QCD used in \cite{Braun:2020ada}, leading to  non-trivial $\beta,\delta$'s. Note also that the extraction of non-trivial scaling is technically not challenging in the fRG approach. In short, non-trivial scaling including sub-leading contributions can hardly be overlooked in fRG computations.

In \cite{Braun:2020ada} it has been shown, that critical scaling is absent (in the approximation used there) at least for pion masses $m_\pi\gtrsim 30$\,MeV. Hence, the results in \cite{Braun:2020ada} do not suggest a large scaling regime. This corroborate previous detailed scaling analyses including the study of finite volume scaling in advanced low energy effective theories of QCD within the past two decades, see in particular \cite{Braun:2010vd} and the review \cite{Klein:2017shl}.

In conclusion, the above mentioned intricacy of the determination of the universality class is of subleading importance for the extraction of the chiral transition temperature in the chiral limit. Owing to the small critical regime and the proximity of the critical scaling to the mean-field one, we arrive at a critical temperature in the chiral limit of
\begin{align}\label{eq:Tc0+error}
	T_{c0}=141.3^{+3.0}_{-3.3}\, \textrm{MeV}\,.
\end{align}
The error estimate in \cref{eq:Tc0+error} is derived from the systematic error estimate \textit{(ii)}, leading to an error of $\pm$5\,MeV for the physical point, see \cref{eq:TcPhys+error}. The critical temperature in the chiral limit, \cref{eq:Tc0+error} agrees quantitatively with the fRG-estimate in~\cite{Braun:2020ada}, while it is consistent with the lattice estimates in~\cite{Ding:2019prx} with $T_{c0}= 132^{+3}_{-6}$\,MeV, and \cite{Kotov:2021rah} with $T_{c0}= 134^{+6}_{-4}$\,MeV. We emphasise that the coincidence of the chiral transition temperature in the chiral limit as well as for all light current quark masses investigated with the fRG results is rather non-trivial in view of the different approximations involved: while the present work includes, in contradistinction to \cite{Braun:2020ada}, the full quark-gluon vertex for all temperatures, the fRG-computation in \cite{Braun:2020ada} includes multi-scattering vertices of the scalar-pseudoscalar four-quark channel ($\sigma$ and pion). The latter allow for a quantitative access to criticality and the size of the scaling regime, which is missing in the present work. In combination, a consistent picture emerges: the scaling regime is rather small, and the largest systematic error left within functional approaches concerns the slope of the chiral transition temperature $T_c(H)$.

\renewcommand{\arraystretch}{1.5}
\begin{table*}[t]
	\centering
	\begin{tabular}{|c|c|C{0.74cm}|C{0.74cm}|C{0.74cm}|C{0.74cm}|C{1.1cm}|C{0.74cm}|C{1.1cm}|C{0.74cm}|C{1.1cm}|C{0.74cm}|C{1.1cm}|C{0.85cm}|C{0.85cm}|C{0.85cm}|}
		\hline
		\multicolumn{2}{|>{}c|}{}&$D(\pi)$ &\multicolumn{12}{c|}{$m_{\pi}$ [MeV]}\\
		\arrayrulecolor{kugray5}
		\arrayrulecolor{black}
		\cline{4-15}
		\multicolumn{2}{|>{}c|}{} && 0& $30$ & $40$ & $55$ & $70$ & $80$ & $100$ & $110$ & $120$ &  $140$& $210$&$370$\\
		\hline
		\multirow{6}{*}{$T_c$[MeV]} & fRG-DSE&0.101 &141.3& $144.2$ & $145.3$ & $146.5$ & $148.3$ & $149.1$ & $151.3$ & $152.1$  & $153.2$ & $155.4$& $161.4$ & $177.4$  \\
		\cline{2-15}
		& fRG&0.10 &-- & $145.3$ & $146.4$ & $148.0$ & $149.6$ & $150.5$ & $152.7$ & $153.6$  & $154.8$ & $156.3$&-- & -- \\
		\cmidrule[1.pt]{2-15} 	
		&hotQCD: $N_{\tau}$ = 8 \hspace{.2cm}
		&0.106&--  & -- & -- & $150.9(4)$ & -- & $153.9(3)$ & -- & $157.9(3)$ & -- & $161.0(1)$ &--&-- \\
		\cline{2-15}
		&
		hotQCD: $N_{\tau}$ = 12 \hspace{.1cm}
		&--&-- & -- & -- & -- & -- & $149.7(3)$ & -- & $155.6(6)$ & -- & $158.2(5)$ &--&--\\
		\cline{2-15}
		&hotQCD: Cont.~Ex.\hfill
		&0.189&132  & -- & -- &-- & -- & $145.6(4)$  & -- & $151.1(6)$ & -- & $157(2)$ &--&-- \\
		\cline{2-15}
		&KLT: fixed scale
		&0.143&--  & -- & -- & -- & -- &-- & -- & -- & -- & $157.8(7)$ & $172(3)$& $197(2)$ \\
		\hline
	\end{tabular}
	\caption{crossover temperatures $T_c$, defined as the peak position of the reduced susceptibility, \cref{eq:Tc} with $\chi_M^{(l,s)}$, for various pion masses from the present computation (fRG-DSE) in comparison with results from~\cite{Braun:2020ada} (fRG), and recent lattice QCD studies: \cite{Ding:2019prx} (hotQCD) with $N_\tau =8,12$ and \cite{Kotov:2021rah} (KLT) with a fixed scaled approach. These results are also depicted in \Cref{fig:Comp}. }
	\label{tab:tc}
\end{table*}
%

\subsection{Combined error estimate for the chiral transition temperature}\label{sec:Constraints}

This situation calls for a combined functional-lattice study: the chiral limit (and the continuum limit) are difficult to access within lattice computations, while functional computations are well-controlled in the chiral limit. However, lattice computations provide benchmark results for current quark masses, where the continuum extrapolation can safely be done. Given the simple dependence of $T_c$ on the current quark mass, as predicted by functional approaches, a  specifically important benchmark result is that of the  slope of $T_c(H)$ in this regime. This goes beyond the scope of the present work, here we only present a preliminary discussion on the basis of the existing functional and lattice data.

To begin with, we list the data of the current computation  in comparison with the fRG-results, \cite{Braun:2020ada}, and lattice results, \cite{Ding:2019prx, Kotov:2021rah} in \Tab{tab:tc}. Given the long regime with linear dependence of $T_c(m_\pi)$ on the pion mass (or a square root dependence on the light current quark mass) discussed in the last section, we consider
\begin{align}\label{eq:Dmpi}
	D(m_\pi)=\frac{T_c(m_\pi)-T_{c0}}{T_{c0}}\,,
\end{align}
the weighted difference of the transition temperatures in the chiral limit and at the physical point. In \cref{eq:Dmpi} the errors stems from the systematic error estimate related to \textit{(ii)} discussed in \Cref{sec:SysError}. For the sake of a direct comparison to the fRG and lattice results we should use the transition temperature for $m_{\pi}= 140$\,MeV. In the present work, the respective critical temperature is given by
\begin{align}\label{eq:Tc140+error}
	T_{c}(m_\pi=140\,\textrm{MeV})=155.4^{+5.1}_{-5.5} \textrm{MeV}\,.
\end{align}
see also \Cref{tab:tc}. The error estimate in \cref{eq:Tc140+error} is derived from the systematic error estimate \textit{(ii)}, leading to an error of $\pm$5\,MeV for the physical point, see \cref{eq:TcPhys+error}. With \cref{eq:Tc140+error} and \cref{eq:Tc0+error} we get
\begin{align}\label{eq:FunSlope}
D(m_{\pi}=140\,\textrm{MeV})=0.101^{+0.013}_{-0.018}\,,
\end{align}
where the error estimate derives from that of $T_{c0}$ in \cref{eq:Tc0+error} and \cref{eq:Tc140+error}. Note also that the ratio grows with increasing coupling strength of the quark-gluon vertex.

\begin{figure}[b]
	\includegraphics[width=.95\columnwidth]{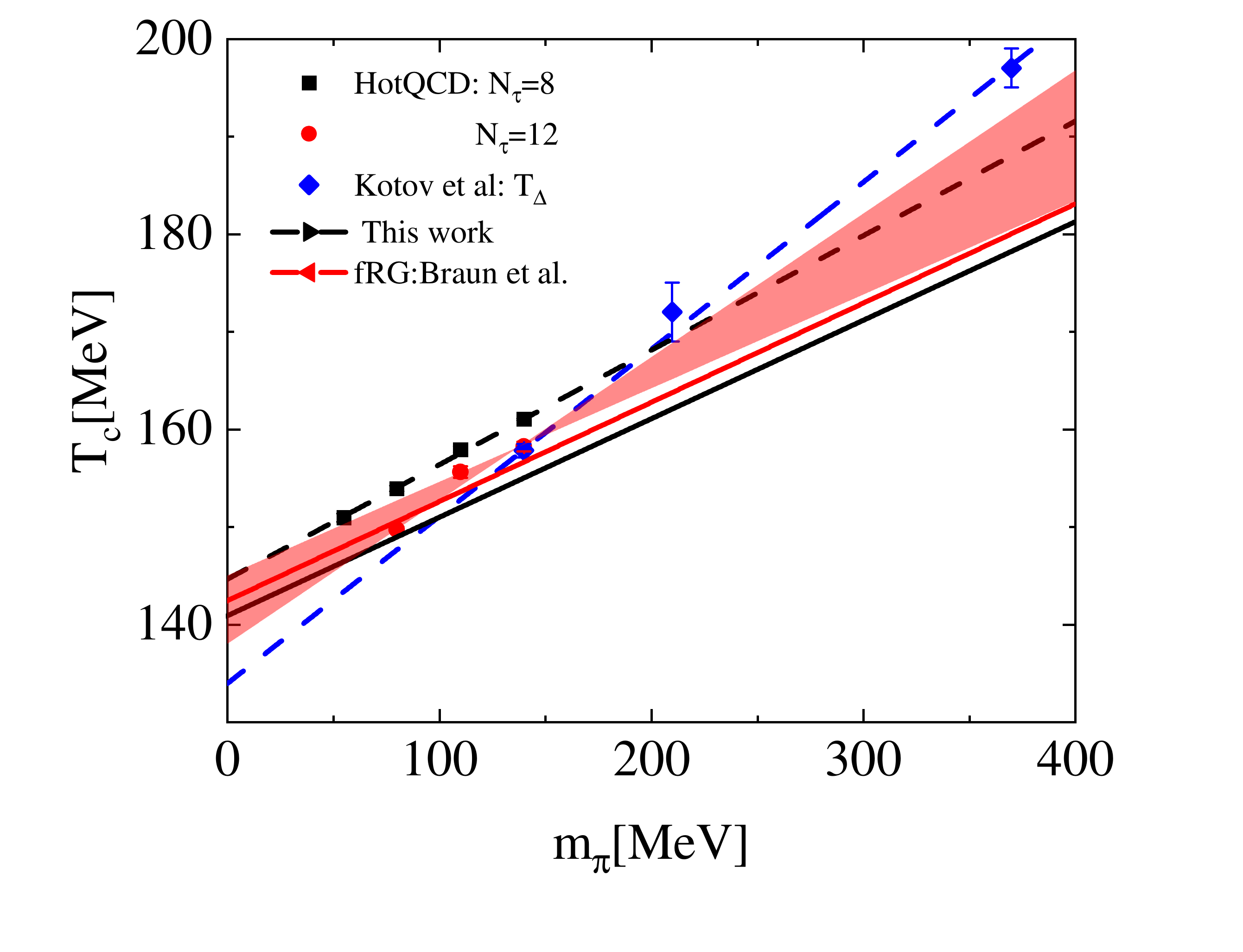}
	\caption{Linear extrapolation of $T_c$ as a function of $M_\pi$ with the lattice data from \cite{Ding:2019prx} (hotQCD) and \cite{Kotov:2021rah} (Kotov et al) and the present functional results and that of \cite{Braun:2020ada} (fRG: Braun et al). The red area indicates extrapolations of the $N_\tau=12$ results with or without the data point at $m_\pi=80$\,MeV. }\label{fig:Comp}
\end{figure}
The result \cref{eq:FunSlope}, obtained in the present DSE approach has to be compared with the respective predictions from the fRG approach in \cite{Braun:2020ada} and lattice predictions \cite{Ding:2019prx, Kotov:2021rah} in \Cref{tab:tc}. While the functional results are in quantitative agreement, the lattice results and functional results are still compatible, if one takes into account an enlarged systematic error for the continuum and infinite volume extrapolation on the lattice as that indicated in \Cref{tab:tc}. In our opinion, such an enlarged systematic error follows straightforwardly from the very  thorough studies of the finite volume extrapolation in low energy effective theories in \cite{Braun:2005fj, Braun:2010vd, Klein:2017shl}. Interestingly, these works indicate a rather strong  flattening of the ratio $D(m_\pi)$, \cref{eq:Dmpi}, in the infinite volume limit. Moreover, whether or not such a strong effect is also present in the continuum limit is difficult to assess, leading to a conservative extension of the systematic error in comparison to that shown in  in \Cref{tab:tc}.

However, as discussed before, the large systematic error can be crunched down in the combined analysis with functional approaches and lattice simulations: The situations suggests a precise lattice determination of the slope within a current quark mass regime, in which the continuum limit can be taken safely. Moreover, this regime should be within the regime of validity of the square root dependence $T_c(m_l) \propto \sqrt{H}$. For these current quark masses chiral expansions such as chiral perturbation theory hold, which is discussed in the next section.

\section{Quasi-massless modes and chiral expansions}\label{sec:ChiralMassless}
	
The results for the magnetic equation of state and the deduced chiral phase structure presented in the last section and depicted in \Cref{fig:Tc} and  \Cref{fig:TcMpi} await an explanation of the underlying dynamics. In particular we expect the occurrence of quasi-massless modes as well as (partial) chiral symmetry (e.g.\ $O(4)$ or $O(2)$)  close to the chiral transition line for sufficiently low current quark masses. The presence of the latter is relevant for the reliability of chiral models used for heavy ion phenomenology, and in particular for chiral transport, e.g.~\cite{Bluhm:2018qkf, Bluhm:2020rha, Grossi:2021gqi, Florio:2021jlx}. Trivially, massless modes and chiral symmetry is present in the critical scaling regime close the the chiral limit. This has spurred many studies of the size of the critical region, and we discuss our present findings in \Cref{sec:SizeCrit} in comparison to that in the literature. In short, the present work is compatible with the findings in previous functional studies that hint at a very small critical region. However, we emphasise that the validity of these chiral transport models or chiral fluid dynamics is not restricted to the critical region, they simply require quasi-massless modes and (partial) chiral symmetry. These properties may be present even far outside the critical region, and are tied to the chiral dynamics.

The details of the chiral dynamics are conveniently accessed with the order parameter potential, $V_\textrm{eff}(\Delta_l)$. Higher order terms corresponded to multi-quark scattering processes in the scalar interaction channel, see in particular \cite{Fu:2019hdw}. In \Cref{sec:OrderPot} we derive the order parameter potential in the DSE approach for the first time, and discuss the relation of these processes for the scaling of the chiral condensate as well as that of the transition temperature with the light current quark mass. In \Cref{sec:Quasi-Massless} the potential different scaling regimes are then derived from the quark gap equation as well as the magnetic susceptibility.

Finally, in \Cref{sec:QuasiMasslessResults} we use our results on the temperature and current quark mass dependence of the chiral  condensate as well as that on the effective order parameter potential to estimate the size of the chiral regime with quasi-massless modes, and the sub-regimes with different chiral dynamics.

\subsection{Critical scaling and the size of the critical region}\label{sec:SizeCrit}

The size of the regime with critical scaling and the respective universality class have been discussed at length in the literature, see e.g.~\cite{Braun:2020ada, Ding:2019prx, Kotov:2021rah} . In particular it has been argued in~\cite{Braun:2020ada}, that the scaling window for $m_\pi\to 0$ in 2+1 flavour QCD is very small and critical scaling is not visible for $m_\pi\gtrsim 30$\,MeV. This finding in functional QCD is supported by respective findings in low energy effective theories, see e.g.~\cite{Braun:2007td, Braun:2009ruy, Braun:2010vd} and the review \cite{Klein:2017shl}. In the latter works it is also shown that apparent critical scaling can be found in regimes far away from the scaling window: These regimes do not show critical scaling, but it may require a high accuracy of the numerical data for extracting its absence.

While this intricate situation has only been demonstrated explicitly for low energy effective theories, the respective theories carry the dynamics of the critical modes in QCD. In terms of scaling, QCD should be seen as a driven model where the driving terms are provided by the glue dynamics. The latter dynamics does not trigger chiral criticality as it carries no chiral symmetry. Typically, the scaling windows of driven models with a non-critical driving force shrink in comparison to the model in the absence of driving forces.  In conclusion the above findings strongly suggest a very small scaling window in QCD for pion masses $m_\pi \ll m_{\pi,\textrm{phys}}$.

The present DSE study cannot add much to this specific question as the current approximation does not include the full dynamics of the chiral critical modes in contradistinction to the fRG study in~\cite{Braun:2020ada}. However, as shown in \Cref{sec:MagEoS} and \Cref{sec:Constraints}, the present results agree quantitatively with that of~\cite{Braun:2020ada}, which corroborates the irrelevance of critical scaling for the magnetic EoS, the transition temperature $T_c(H)$ and in particular for $T_{c0}$. Explicitly, we consider the finite volume expansion in the 3d\,$O(4)$ universality class as used e.g.~in \cite{Ding:2019prx},
\begin{align}
	\label{eq:expan}
	T_p(H,L)=T_{c0}\left(1+\frac{z_X(z_L)}{z_0}H^{\frac{1}{\beta\delta}}\right)+c_X H^{1-\frac1\delta+\frac{1}{\beta\delta}},
\end{align}
where $L$ is the size of the system, and $L\rightarrow\infty$ in the continuum limit. \Cref{eq:expan} also holds for other universality classes, and it requires the computation of the two critical exponents $\beta$ and $\delta$. The former one can be derived from the temperature-dependence of the chiral condensate in the chiral limit in the critical region with $T\to T_{c0}$ from below,
\begin{align}\label{eq:beta}
	\Delta_{l,\chi}(T)  \propto \left( \frac{T_{c0}-T}{T_{c0}}\right)^\beta\,,\quad \beta = \frac12 \nu (d-2 +\eta_\Delta)\,,
\end{align}
with the spacial dimension $d$. The critical exponent $\delta$ describes the scaling with the light current quark mass in the critical region with $H\to 0$ and $T=T_{c0}$,
\begin{align}\label{eq:delta}
\Delta_{l}(T_{c0},H) \propto H^{\frac{1}{\delta}}\,,\qquad \delta= \frac{2+d-\eta_\Delta }{d-2+\eta_\Delta}\,.
\end{align}
In \Cref{sec:Quasi-Massless} we will derive (non-critical) scaling laws for the chiral condensates in order to determine the size of the regime with massless modes rather than that with critical scaling. In any case, for the present purpose of critical scaling the asymptotic cases are,
\begin{enumerate}
\item[(1)] Asymptotically large critical region: Then all relevant pion masses are covered by the scaling regime, and we can globally use \cref{eq:expan}. In case of 3d\,$O(4)$ scaling, the scaling exponents are given by $\beta=0.379$, $\delta=4.820$ and $1/(\beta\delta)=0.547$. For comparison, the 3d\,$O(2)$ scaling exponents are given by $\beta=0.349$, $\delta=4.780$  and $1/(\beta\delta)=0.599$, the 3d\,$Z(2)$ scaling exponents of a potential CEP are given by $\beta=0.326$, $\delta=4.805$ and $1/(\beta\delta)=0.638$.
\item[(2)] Asymptotically small critical regime with 3d\,$O(4)$ scaling: This implies that the order parameter potential has a polynomial expansion for most parameter values. We shall see that
\begin{align}
	\label{eq:subleading}
	T_p(H,L)=T_{c0}\left(1+a(L)H^{\frac12}+b(L)H\right)\,,
\end{align}
fits the data well even in the chiral limit $m_l=0$ with $m_\pi=0$. The $H$-scaling of $T_p$ is linked to that of $\Delta_l$ in the presence of quasi-massless modes, which is discussed in detail in the next sections, \Cref{sec:OrderPot} and \Cref{sec:Quasi-Massless}.
\end{enumerate}
Now we use our data to extrapolate to $T_{c0}$ assuming either case {(1)} with 3d\,$O(4)$ scaling or case (2). We found that both extrapolations lead to $T_{c0}\approx 141$\,MeV. Similar results are achieved for $O(2)$ or $Z(2)$ critical exponents. Indeed, our data are matched best with \cref{eq:subleading}, and the lattice data are also well compatible with this trivial relation that follows from a simple chiral expansion. We also note that a possible volume dependence of $c_X$ and further volume dependences largely add to a systematic error estimate, as discussed at the end of \Cref{sec:Constraints}, see also \Cref{fig:Comp}. This leads us to a conservative combined estimate for $T_{c0}$,
\begin{align}\label{eq:CombinedEstimate}
132\,\textrm{MeV} \lesssim T_{c0} \lesssim  141\,\textrm{MeV}\,,
\end{align}
where the range is solely due to the combined systematic error estimate based on functional results from the present DSE approach  and from the fRG computation in \cite{Braun:2020ada}, as well as lattice results from \cite{Ding:2019prx, Kotov:2021rah}. While the two approaches also allow for smaller (lattice QCD) or larger (functional QCD) temperatures within the respective error estimates, in combination these temperatures are disfavoured.

In summary this situation calls for a combined study  with functional and lattice approaches for crunching down the relatively large systematic error in \cref{eq:CombinedEstimate}, exploiting the respective strengths. This task is left to future work.

\subsection{Order parameter potential}\label{sec:OrderPot}

All scalings, and in particular the critical scaling discussed in the last Section, can be conveniently extracted from the temperature-dependent order parameter potential $V_\textrm{eff}(\Delta_l,m_l)$ at fixed $m_s$. It also offers a more direct access to the related physics: The effective potential of the order parameter $\Delta_{l}$ can be derived from the effective action $\Gamma[A, c,\bar c, q,\bar q; \Delta_l, m_l]$ at fixed $m_s$, where we have introduced a current for $\Delta_l$ and applied a respective Legendre transform. In the fRG approach this is done with \textit{dynamical hadronisation}, more details in the present context can be found in \cite{Fu:2019hdw, Braun:2020ada, Dupuis:2020fhh}. Then, the order parameter potential is given by a convex function $V_\textrm{eff}(\Delta_l,m_l)=\Gamma[0;\Delta_l,m_l]/$(volume). Moreover, it can be shown that the only term with the light current quark mass is linear in $m_l$. For a recent detailed discussion of this fact see \cite{Fu:2019hdw}. We have
\begin{align}\label{eq:Veff}
V_\textrm{eff}(\Delta_l,m_l) = V_\chi(\Delta_l) - c_\Delta \,m_l\,\Delta_l\,,
\end{align}
with a dimensionless constant $c_\Delta$. While such a split very much resembles chiral expansions, it is important to note that \cref{eq:Veff} is valid for all masses. The temperature and current-quark mass dependent expectation value of the condensate is determined by the solution of the equation of motion (EoM) $\Delta_\textrm{EoM}$,
\begin{align}
	\label{eq:DeltaEoM}
\left. \frac{\partial V_\chi(\Delta_l)}{\partial\Delta_l}\right|_{\Delta_l=\Delta_{l,\textrm{EoM}}(T,m_l)} = c_\Delta \,m_l\,.
\end{align}
Hence, \cref{eq:Veff} carries the thermal and current mass dependence of the chiral condensate, and can be used to extract the transition temperature. The potential $\bar V_\chi$ gives us easy access to the scaling behaviour of the pion mass. On the EoM for $\Delta_l$ we have,
\begin{align}
	m_\pi^2\propto \frac{\partial V_\chi(\Delta_l)}{\partial \Delta_l^2}=  \frac{c_\Delta m_l}{2\Delta_{l,\textrm{EoM}}(T,m_l)}\,,
\label{eq:mpi}\end{align}
which also allows us to recover $\chi$PT in the vacuum, for a detailed discussion see e.g.\ \cite{Fu:2019hdw}. The effective potential also carries the (quark) pressure with $V_\chi(\Delta_\textrm{EoM}) - c_\Delta m_l\,\Delta_\textrm{EoM}$, which has been used in \cite{Isserstedt:2020qll} for a determination of the EoS. The temperature-independent normalisation $c_\Delta$ can be conveniently fixed at large temperatures, where the pressure resumes the Stefan-Boltzmann limit. However, in the present work we are not interested in the latter, and for the sake of simplicity of the following considerations we normalise \cref{eq:Veff} with $c_\Delta$,
\begin{align}\label{eq:barVeff}
	\bar V_\textrm{eff}(\Delta_l,m_l) = \bar V_\chi(\Delta_l) -  m_l\,\Delta_l\,, \end{align}
with $\bar V_\chi(\Delta_l)=V_\chi(\Delta_l)/c_\Delta$ is the potential in the chiral limit, divided by the dimensionless constant $c_\Delta$. Formulated in terms of $\bar V_\chi$, the EoM in \cref{eq:DeltaEoM} turns into
\begin{align}
	\label{eq:barDeltaEoM}
\frac{\partial \bar V_\chi(\Delta_l)}{\partial\Delta_l}= m_l(\Delta_l)\,.
\end{align}
\Cref{eq:barDeltaEoM} has already been adapted to our present results: we have computed $\Delta_l(T,m_l)$ for a given $H=m_l$, which provides us with pairs of $(\Delta_l(T), m_l(T))$ for all temperatures. Accordingly, the chiral effective potential $V_\chi(\Delta_l,T)$ for $\Delta_l(T)>\Delta_{l,\chi}(T)$ can be represented as
\begin{figure}[t]
	\hspace{-.5cm}	\includegraphics[width=1\columnwidth]{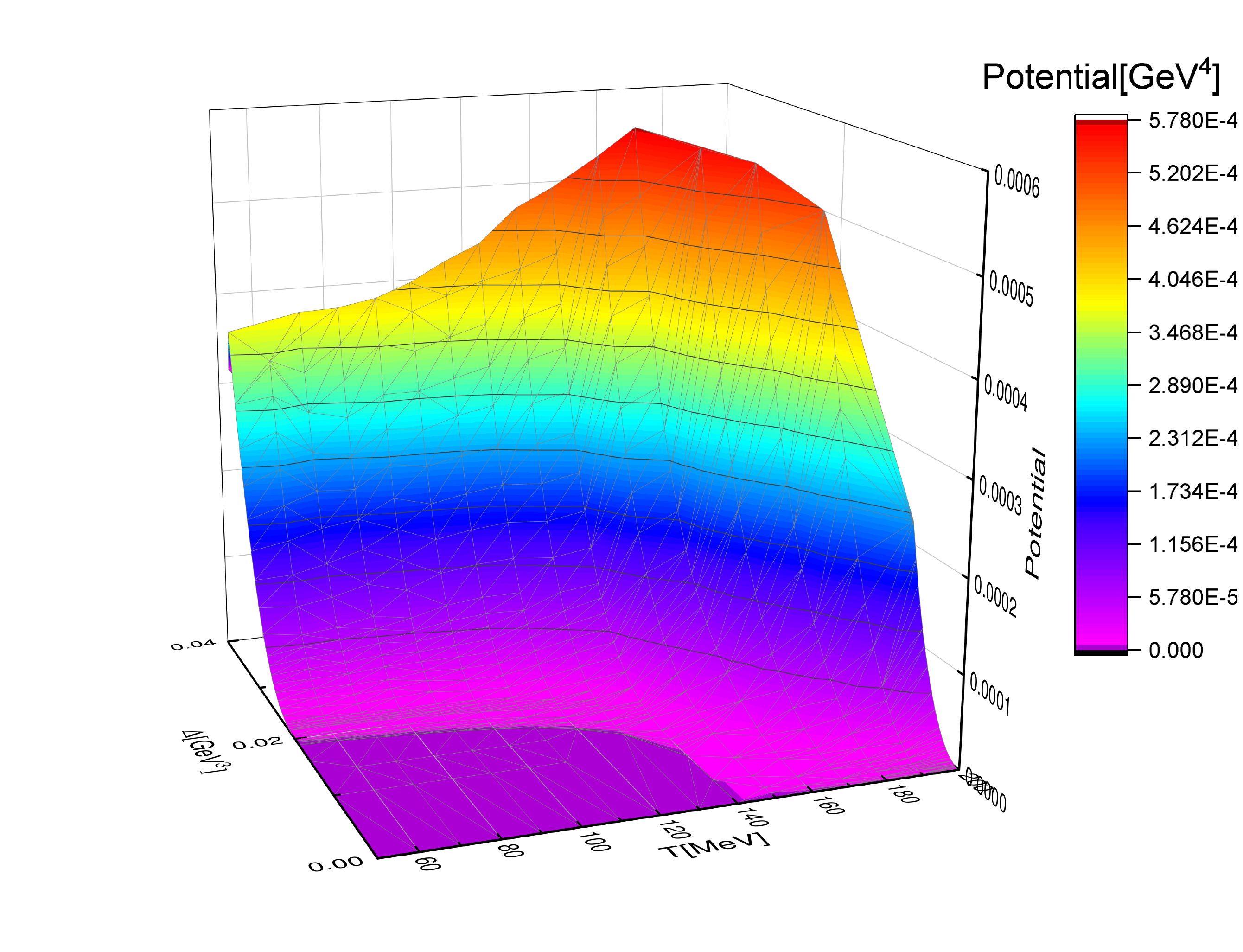}
	\caption{Chiral effective potential  $V_\chi(\Delta_l)$ as a function of the chiral condensate $\Delta_l$ and temperature $T$. The boundary of the flat regime is the solution of the EoM in the chiral limit. $V_\chi(\Delta_l)$ is fit well with the polynomial potential \cref{eq:Potbreaking} (broken regime) and \cref{eq:Potsym} (symmetric regime), the respective couplings are depicted in \Cref{fig:lambdai}.}\label{fig:EffPot}
\end{figure}
\begin{subequations} \label{eq:Vcomp}
\begin{align}\label{eq:VcompDelta}
\bar V_\chi(\Delta_l,T) = \int_{\Delta_\chi(T)}^{\Delta_l} d \Delta\,  m_l(\Delta,T) \,.
\end{align}
For $\Delta_l< \Delta_{i,\chi}$ we have $\bar V_\textrm{eff}(\Delta_l,T) = \bar V_\textrm{eff}(\Delta_{l,\chi},T)$ due to convexity of $V_\chi$. The $\Delta$-integral in \cref{eq:VcompDelta} can be turned into an $H$-integral with the help of the chiral susceptibility
\cref{eq:MagSuscept},
\begin{align}\label{eq:VcompH}
\bar V_\chi(\Delta_l,T) =  - \frac12 \int_0^{m_l^2(\Delta_l)} d m^2_l\,\chi^{(l)}(m_l,T) \,,
\end{align}
\end{subequations}
for $\Delta_l>0$. In the presence of a first order phase transition, \cref{eq:Vcomp} has to be slightly modified, but a similar expression still holds true. We also note that the potential is even in $\Delta_l\to -\Delta_l$ which reflects the symmetry under $m_l\to -m_l$.

This leaves us with the following task: For a given temperature $T$ we compute numerically the $m_l$- or $H$-dependence of the light chiral condensate $\Delta_l$, leading to a fit of $\Delta_l(T,H)$. The integral representation  \cref{eq:Vcomp} of the chiral effective potential requires $m_l(T,\Delta_l)$ or $H(T,\Delta_l)$, which is extracted by numerically inverting this relation. The result is depicted in \Cref{fig:EffPot}. The flat region for $\Delta_l\leq \Delta_{l,\chi}(T)$ shrinks with temperature and vanishes for $T\geq T_{c0}$. Moreover, we clearly see the flattening of the potential in the vicinity of $T_c$ for $\Delta_l>\Delta_{l,\chi}(T)$, indicating the emergence of quasi-massless modes.

\begin{figure}[t]
	\includegraphics[width=.93\columnwidth]{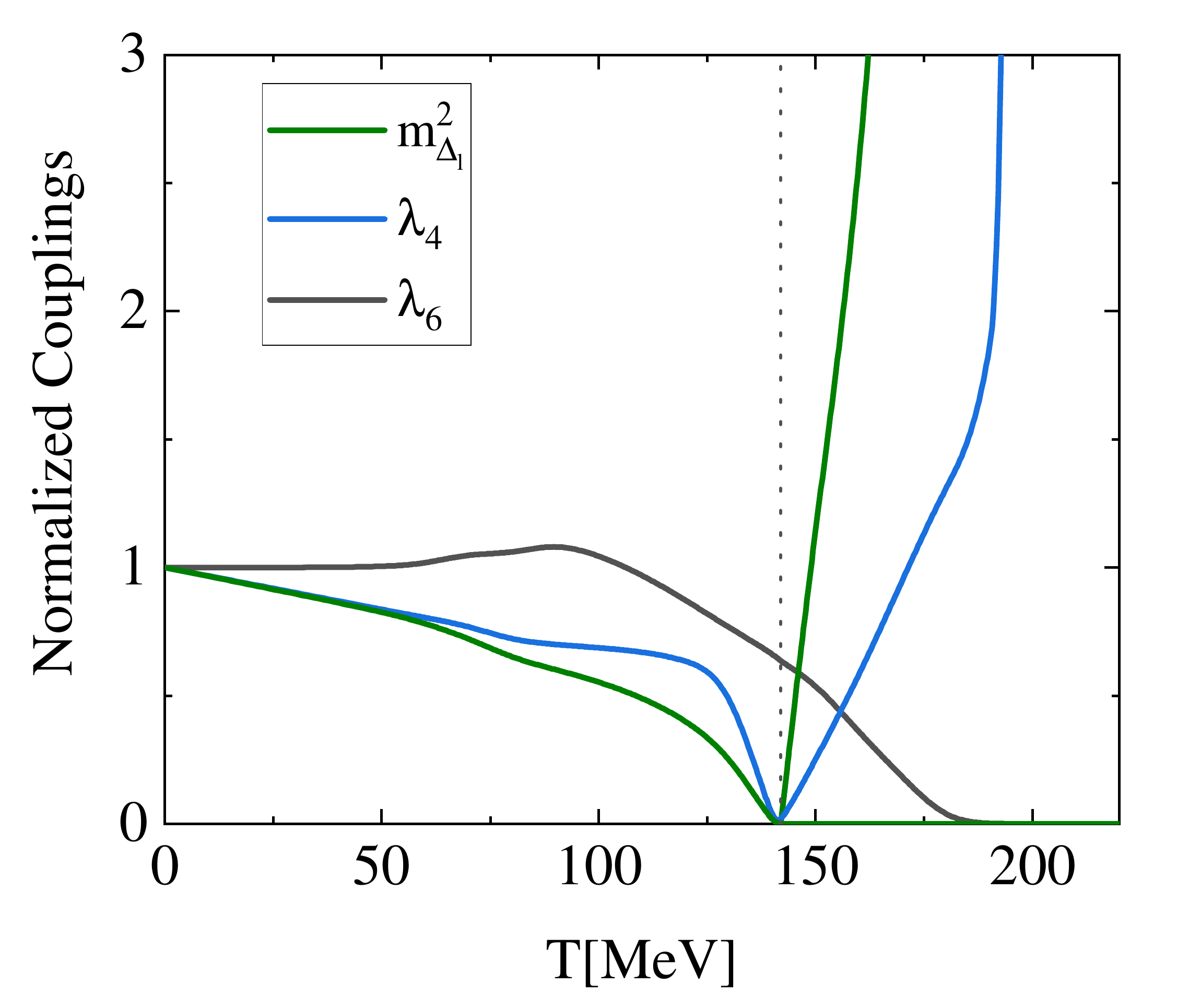}
	\caption{Temperature dependence of the mass and coupling parameters in the polynomial potentials \cref{eq:Potbreaking} with \cref{eq:mDeltabroken} (broken regime) and \cref{eq:Potsym} (symmetric regime). The couplings have been normalised by their values at $T=0$. At vanishing temperature, $T=0$, we have  $m_\Delta^2=0.24$\,GeV$^2$, $\lambda_4=3.3\times10^3$, $\lambda_6 =3\times10^6$\,GeV$^{-2}$. }\label{fig:lambdai}
\end{figure}
We also can provide simple polynomial fits for the effective potential shown in \Cref{fig:EffPot}: In the broken regime we consider a polynomial potential,
\begin{align}\label{eq:Potbreaking}
	V_\chi(\Delta_l) = \frac{\lambda_4}{8}(\Delta_l^2- \Delta_{l,\chi}^2)^2+ \frac{\lambda_6}{24}  (\Delta_l^2- \Delta_{l,\chi}^2)^3\,,
\end{align}
where light chiral condensate in the chiral limit, $\Delta_{l,\chi}\geq 0$, minimises the potential. \Cref{eq:Potbreaking} is the potential of the scalar ($\sigma$-) mode, and the respective 'mass' in the broken regime is
\begin{align}
	\label{eq:mDeltabroken}
	m_{\Delta_l}^2 = \partial_{\Delta_l}^2 V_\chi(\Delta_l=\Delta_{l,\chi})=\lambda_4\, \Delta_{l,\chi}^2\,.
\end{align}
Hence, \cref{eq:Potbreaking} provides a three-parameter fit to the potential in the broken regime in an expansion about the minimum $\Delta_{l,\chi}>0$. The parameters are $\Delta_{l,\chi}, \lambda_4, \Lambda_6$, the mass at the minimum is a derived quantity. However, with \cref{eq:mDeltabroken}  we can also use $m_{\Delta_l}^2$ as parameter instead of $\Delta_{l,\chi}^2 =  m_{\Delta_l}^2/\lambda_4$, leaving us with the three parameters $m_{\Delta_l}^2, \lambda_4, \Lambda_6$.

\begin{figure}[t]
	\includegraphics[width=.95\columnwidth]{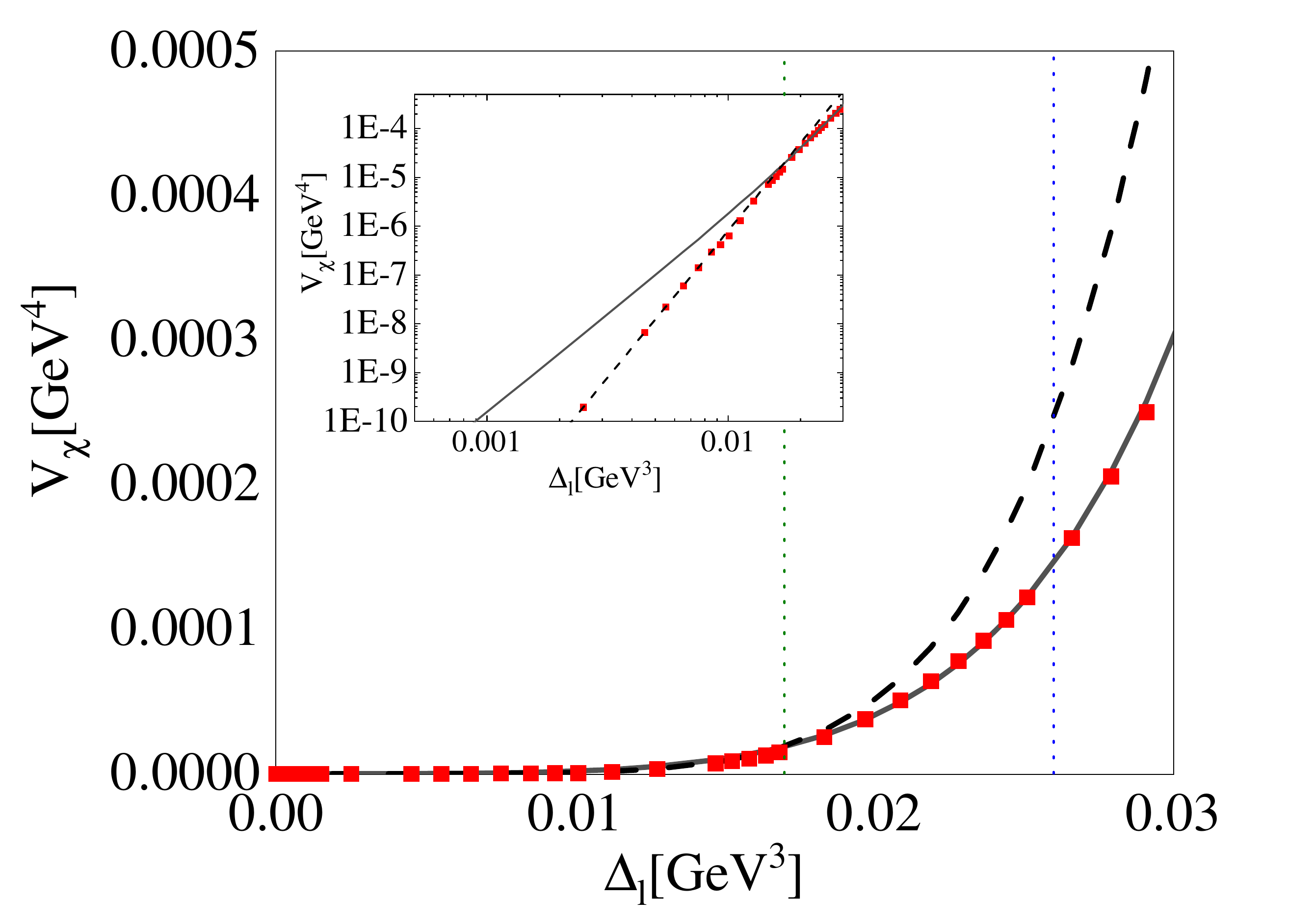}
	\caption{Order parameter potential $V_\chi(\Delta_l)$ (\textit{red squares}) for the chiral phase transition temperature in the chiral limit, $T_{c0}$. The potential is shown for the regime \cref{eq:ChExp} with $\Delta_\textrm{ch}=0.026$\,GeV$^3$. We also display the fits computed from the regime \cref{eq:fitchiral}
		with $\Delta_{\textrm{max}}=0.017$\,GeV$^3$ (\emph{dashed curve}), as well as that computed from the regime $\Delta_l\in [\Delta_{\textrm{max}}, \Delta_\textrm{ch}]$ (\emph{solid curve}). Outside the regime \cref{eq:ChExp} the latter fit fails. }\label{fig:VTchi}
\end{figure}
In the symmetric regime we have the fixed expansion point $\Delta_{l,\chi}=0$, and the expansion point parameter is in any case traded for the explicit mass parameter. Again this leaves us with the three parameters, $m_{\Delta_l}^2, \lambda_4, \Lambda_6$, and the the potential is given by,
\begin{align}\label{eq:Potsym}
	V_\chi(\Delta_l) =  \frac{	m_{\Delta_l}^2}{2}\Delta_l^2+\frac{\lambda_4}{8}\Delta_l^4+ \frac{\lambda_6}{24}  \Delta_l^6\,.
\end{align}
Both, in the broken and in the symmetric regime we have a Gau\ss ian mass term, and and a standard $\Delta_l^4$ interaction term with the interaction strength $\lambda_4$. The $\Delta_l^6$-term carries higher order interactions of the scalar interaction channel of quark--anti-quark scatterings. The parameters are now obtained in a fit regime about the respective minimum $\Delta_{l,\chi}$, which is one of the parameters in the broken regime. We use a $\chi^2$-fit in a fit regime
\begin{align}
	\label{eq:fitchiral}
\Delta_l\in[\Delta_{l,\chi}(T), \Delta_\textrm{max}(T)]\,.
\end{align}
In \cref{eq:fitchiral}, $\Delta_\textrm{max}$ is the maximal $\Delta_l$, for which the fit parameters as well as the $\chi^2$ is stable. In the vicinity of $T_{c0}$ the respective $\Delta_\textrm{max}$ is shrinking, and we also use additional polynomial fits in the regime $\in[\Delta_\textrm{max}, \Delta_\textrm{ch}]$ for an estimate of the regime dominated by the  $\Delta_l^4$ term. It is suggestive to speculate that chiral expansions work in the combination of these regimes, that is the regime
\begin{align}{\label{eq:ChExp}}
\Delta_l\in[\Delta_{l,\chi}(T),  \Delta_\textrm{ch}(T)]\,.
\end{align}
Note however, that the regime \cref{eq:ChExp} does not sustain a global fit as close to $T_{c0}$ the potential is dominated by a higher order polynomial.

The temperature dependence of the three parameters, normalised by their value at vanishing temperature, is depicted in  \Cref{fig:lambdai}. The steep drop of both, $m_{\Delta_l}^2(T)/m_{\Delta_l}^2(0)$ and $\lambda_4(T)/\lambda_4(0)$ for $T\to T_{c0}$ from below signals the approach to the chiral phase transition. The drop of $\lambda_4$ entails that the $\lambda_6$-term dominates, leading to an $H^{1/5}$ scaling, as discussed below. Note however, that this does not necessarily entail critical scaling in this regime. In \Cref{fig:VTchi} we show the full potential at the chiral transition temperature in the chiral limit, $T_{c0}$, (red squares) together with the monomial fit with $\Delta_l^6$ in the regime \cref{eq:fitchiral} with $\Delta_\textrm{max}(T_{c0})=0.017$\,GeV$^3$ as well as a polynomial fit in the regime $\Delta_l\in[\Delta_\textrm{max}, \Delta_\textrm{ch}]$ with $\Delta_\textrm{ch}=0.026$ GeV$^3$. For chiral condensates larger than $\Delta_\textrm{ch}$ we approach the regime of asymptotically large current quark masses.

We proceed with discussing different regimes in the $T, H$-plane with different chiral or non-chiral dynamics. This dynamics is carried by the interaction terms in the effective potential $V_\chi(\Delta_l)$. We note that for (fractional) monomial scaling of $\Delta_l$ with $H$ the respective term in the effective action can be deduced analytically. This gives us an analytic understanding of the $H$-scaling of $\Delta_L$ and the underlying QCD dynamics. For example, monomial potentials lead to specific monomial $H$-scalings,
\begin{align}\label{eq:VnScaling}
	V_n(\Delta_l)\approx  \frac{\lambda_n}{n!} (\Delta_l^2-\Delta^2_{l,\chi})^{\frac{n}{2}}\,,\quad \to \quad  \Delta_l(H) \propto H^{\frac{1}{n-1}}\,,
\end{align}
for large $H$. \Cref{eq:VnScaling} provides the link of the $H$-dependence of the condensate for different temperature with the dynamics of the chiral condensate as $\Delta_{l}^n$ terms in the effective potential simply entail the multi-scattering of scalar quark-antiquark channels. This relation is directly seen within dynamical hadronisation, see e.g.\ \cite{Fu:2019hdw, Braun:2020ada}.

The inversion property $\Delta^n\to H^{1/(n-1)}$ displayed in \cref{eq:VnScaling} entail that higher order terms in the potential lead to terms that dominate an expansion of the chiral condensate for small $H$. Consequently one can hope for accurate global fits $\Delta_l(H)$ in the potential validity regime \cref{eq:ChExp} of chiral expansions. This property will be used in \cref{sec:QuasiMasslessResults}.

Moreover, we emphasise that specifically for $n=4,6$ the scaling in \cref{eq:VnScaling} is the  3d mean field scaling, $\delta=3$,  and 3d dimensional scaling, $\delta=5$, respectively. The latter is derived  from  \cref{eq:delta} with a vanishing anomalous dimension, $\eta_\Delta=0$. Given the small anomalous dimensions for 3d $O(4)$, 3d $O(2)$ (and 3d $Z_2$) universality classes, the dimensional scaling is very close to the full universal scalings, and in particular to the $O(4)$-scaling. In the present DSE approach a non-vanishing anomalous dimension $\eta_\Delta$ requires a full resolution of the (scalar-pseudoscalar) four-quark interaction, typically included via pion contributions. We have not taken into account (the thermal part) of the resonant scalar-pseudoscalar four-quark interaction leading to $\eta_\Delta=0$. Note however, that this has been done in \cite{Braun:2020ada} for $m_\pi\geq 30$\,MeV with no sign of criticality. Still, as mentioned before, the present work does not add much to the intricate question of the size of the critical regime. Note also that in the vacuum we expect an $H^{1/3}$-scaling for large, but not asymptotically large H, which we call \textit{trivial} scaling as it is far away from criticality (at asymptotically large $H$ we expect an $H^3$ scaling fixed by dimensionality). We emphasise that this makes it extremely difficult to disentangle \textit{trivial} scaling regimes from \textit{critical} scaling regimes. This entails, that both, the proof of presence as well as that of the absence of critical scaling requires data with exceedingly small statistical error. This has been studied in detail in ~\cite{Braun:2007td, Braun:2009ruy, Klein:2017shl}, advocating both, a small critical regime and the necessity for exceedingly small statistical error.

Indeed, the worst case in this respect is a trivial scaling regime close to $T_{c0}$, that exhibits an $H^{1/5}$-scaling, stemming from a $\Delta_l^6$-potential. As the quadratic term vanishes at $T_{c0}$, $\lambda_2(T\to T_{c0})\to 0$, the presence of this scenario is linked to the fate of $\lambda_4(T\to T_{c0})$. We shall see that also $\lambda_4(T\to T_{c0})\to0$ and we may be left with a \textit{trivial} $H^{1/5}$-scaling that turns into a critical one with $H^{1/(\delta\beta)}$-scaling in the critical regime. How to distinguish this scenario from the one with a large regime with critical scaling, a more refined analysis is required. This analysis is deferred to a forthcoming publication, \cite{fQCDscaling}.

While this intricacy makes it difficult to disentangle \textit{critical} from \textit{trivial} scaling, it is very good news for chiral expansions and extrapolations. In summary, the present analysis gives full access to the phenomenologically relevant validity regime of chiral expansions with quasi-massless modes. It also corroborates the point of view, that for phenomenological applications and experimental signatures in heavy ion collisions the search for critical scaling and its signatures is rather a red herring than a property to look for.

\subsection{Quasi-massless modes}\label{sec:Quasi-Massless}

With these preparations we now can tackle the phenomenologically important question after the regime with quasi-massless modes close to the transition line, already mentioned in the introduction of \Cref{sec:ChiralMassless}. Chiral transport models or chiral fluid dynamics both rely on the existence of \textit{quasi-massless} modes such as the pion or the sigma mode close to the potential critical end point in the QCD phase structure. Certainly critical modes are \textit{quasi-massless} in the critical regime, but this property is more generally connected to the validity regime of chiral expansions.

The most prominent chiral expansion is chiral perturbation theory ($\chi$PT), an expansion about QCD in the chiral limit in the vacuum.
It is a low energy effective theory of QCD, and the expansion coefficients or low energy constants are either determined from first principles QCD computation in the chiral limit or they are fixed by experimental data. We have already used $\chi$PT results for the determination of the pion mass in the present approach, see \cref{eq:fpi+mpi} and below. $\chi$PT is impressively successful in the vacuum as well as small enough temperatures and densities. In $\chi$PT, the chiral condensate has a chiral expansion similar to that in
\cref{eq:fpi+mpi} for the pion mass and pion decay constant. We parametrise,
\begin{subequations}
	\label{eq:ChiralExpansion}
\begin{align}
	\Delta_l(T,H)= &\,\Delta_{l,\chi}(T)\Bigr[ 1+ \left(c_1+c_\textrm{log}\,\log H\right) \,H +O(H^2)\Bigr]\,,
	\label{eq:ChiralExpDeltaT}\end{align}
with temperature-dependent coefficients $c_\textrm{log}(T)$, $c_1(T)$. The estimate \cref{eq:ChiralExpDeltaT} can only hold for temperatures below $T_{c0}$, so at most $T<T_{c0}$.

Note also that \cref{eq:ChiralExpDeltaT} can be reformulated as a chiral expansion such as \cref{eq:fpi+mpi} with the expansion parameter \cref{eq:GMORchiln}. The logarithm in \cref{eq:ChiralExpDeltaT} originates in massless pion loops in the vacuum within the expansion about $H=0$ with $m_\pi=0$. There are higher order logarithms well-known in chiral perturbation theory, which we discard for the present estimate. Moreover, at sufficiently large temperature we do not expect a logarithmic scaling any more, as the infrared singularities at finite temperature are 3d ones and have a rational scaling. Indeed, we shall see that \cref{eq:ChiralExpDeltaT} does not work for $T\gtrsim 50$\,MeV, see \Cref{fig:ChiralRange}.

Finally, for asymptotically large quark masses the current quark mass dominates the constituent quark mass and hence the condensate. There we expect a trivial linear $m_l$-scaling of the renormalised condensate. Whether this transition already takes place for $H\leq 1$ is a dynamical question, which also depends on the size of the strange quark mass.

In summary, we infer from this discussion, that an $H$-expansion of the chiral condensate for small $H$ and small temperature at least has to incorporate the logarithm and polynomial terms known from $\chi$PT. In turn, for larger $H$ and temperatures the logarithm is absent.

Hence, for sufficiently large temperatures we use a rational expansion without logarithm, as the latter is absent in a three-dimensional chiral expansion,
\begin{align}  \label{eq:Ratxpanse}
	\Delta^{(\chi\textrm{EP})}_l(T,H) \approx &\,\Delta_{l,\chi}(0)\Bigl(c_0 +  c_{r_1} H^{r_1}+ c_{r_2} H^{r_2}+c_1 H\Bigr)\,,
\end{align}
\end{subequations}
for the extension of the $\chi$PT regime to a chiral expansion regime ($\chi$EP) at larger temperatures and small current quark masses. Here, $H^{r_1}$ is the leading scaling for $H\to 0$, and hence $r_1 < r_2$. As in the vacuum, we expect a linear $H$-scaling of the chiral condensate for sufficiently large current quark masses. Then, the explicit chiral symmetry breaking via $H$ dominates over the spontaneous one. This linear regime is expected to set-in at smaller current quark masses for larger temperature, as the chiral condensate in the chiral limit, which measures the amount of spontaneous chiral symmetry breaking, reduces with the temperature.  Hence, \cref{eq:Ratxpanse} may also allow for good global fits, in contradistinction to $\chi$PT, where the logarithmic small $H$-term would also dominate the large $H$-regime.

We proceed by discussing the different $H$-scalings we expect to see from general diagrammatic arguments as well as within the present approximation.

To begin with, for constant vertices the gap equation allow for a simple extraction of mean field scaling for $\Delta_l\to 0$. We shall use that for the purpose of a leading order scaling analysis that the wave functions $Z_q(p)$ and $Z_q^\parallel(p)$ of the quark propagator in \cref{eq:QuarkProp} only show a mild momentum dependence, see e.g.\ the recent DSE review \cite{Fischer:2018sdj}, and \cite{Gao:2020fbl, Gao:2020qsj, Gao:2021wun}, where the DSE framework used in the present work is put forward. Hence, for the present scaling analysis the quark propagator is well approximated by
\begin{align}\label{eq:GqApprox}
	\frac{M_q-\imag\,\slash{\hspace{-.193cm}p}}{p^2+M_q^2}\,,
\end{align}
where $M_q=M_q(p=0)$ is the infrared value of the constituent quark mass. Within this approximation the quark gap equation is an integral equation for $M_q$, and its resolution with a gapped gluon propagator readily leads to
\begin{align}\label{eq:MF}
	M_q(H) \propto H^\frac13 +O(H)\,,
\end{align}
in the limit $H\to 0$. With $\Delta_l \propto M_l$ this entails the mean field scaling for the chiral condensate. It is well-known that the behaviour \cref{eq:MF} is already seen with the crude approximation $G_{AA}(p) \approx \delta(p^2-m_\textrm{gap}^2)$ with $H\ll m_\textrm{gap}$. Note however, that this scaling follows as trivial scaling from a polynomial order parameter potential $V_\chi$ with a (Gau\ss ian) mass term proportional to $\Delta_l^2$ and a $\Delta_l^4$-term (Ginzburg-Landau potential).

\begin{figure}[t] 
	\vspace{.2cm}
	\includegraphics[width=.9\columnwidth]{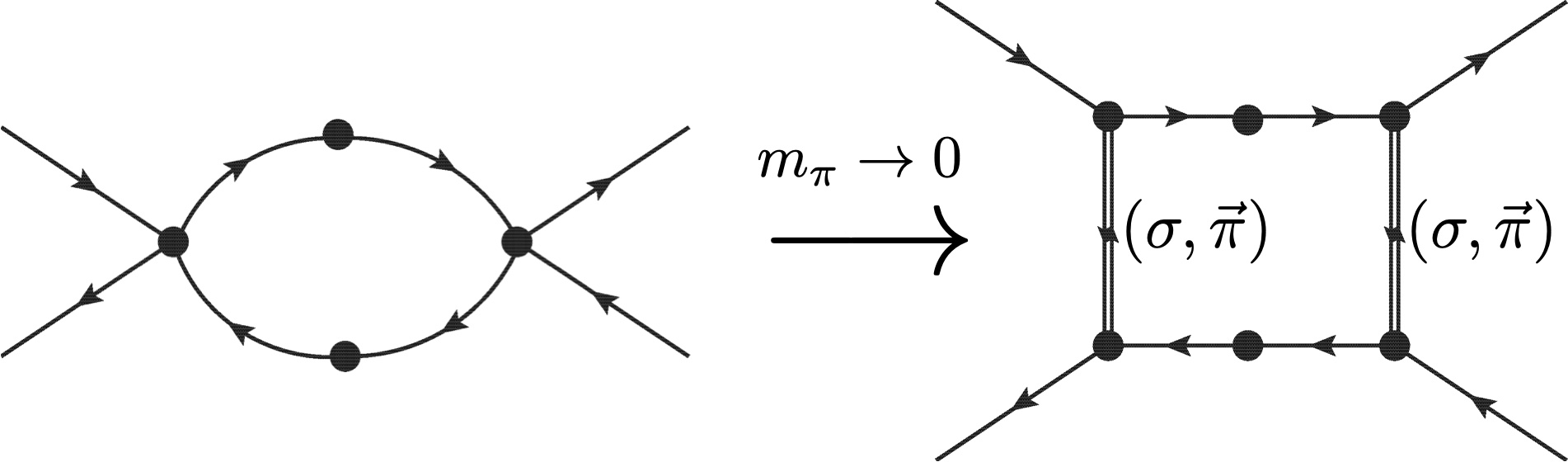}	
	\caption{Leading diagram in the chiral limit, $m_\pi\to 0$ in the diagrammatic representation of the magnetic susceptibility. The left diagram is the (resummed) fish diagram with two dressed four quark vertices, the right diagram shows the dominant scalar--pseudo-scalar $(\sigma,\vec \pi)$-channel in the chiral limit. The pions and the $\sigma$ mode are depicted by a quark double line, indicating its $l \bar l$ content.}\label{fig:FishBox}
\end{figure}
For the full scaling the DSE requires a non-trivial quark-gluon vertex with a full back coupling of fluctuations, including the critical ones we expect the full $H$-scaling to be present in the gap equation. The scaling analysis of this case is conveniently done with the chiral susceptibility: As a current quark mass derivative of the chiral condensate it is an integrated four-quark correlation function. It has a diagrammatic representation in terms of its DSE of fRG equation, which already diagrammatically includes part of the back reaction required for the full scaling. Its chiral limit, $H\to 0$, is governed by the most divergent diagram for $H\to 0$. In the present DSE approach this analysis is best done after introducing effective degrees of freedom for the resonant scalar-pseudo-scalar interaction four-quark channels, as done in the fRG approach with dynamical hadronisation, in the present context see \cite{Fu:2019hdw, Braun:2020ada}. Alternatively one may employ the respective resummations. In the first case this leads us to a quark-pion box diagram, in the latter case to a fish diagram with two dressed four-quark-vertices, see \Cref{fig:FishBox}.

For the sake of a scaling analysis the pseudoscalar resonant four-quark interaction channels (or pion propagators)  may be approximated by $1/(p^2 + m_\pi^2)$ away from the critical scaling regime, and the light and strange quark propagators may still be approximated by \cref{eq:GqApprox}.

Naturally, the analysis of both diagrams leads to the same chiral scaling, and in the vacuum it leads to the well-known results in chiral perturbation theory. At sufficiently large temperature the scaling analysis leads to
\begin{align}\label{eq:T-scaling}
	\frac{\partial \Delta_l}{\partial H}\propto \frac{M_l}{m_\pi (m_\pi+M_l)^3}\,.
\end{align}
\Cref{eq:T-scaling} originates from the 3d spatial momentum integral of the zeroth Matsubara mode.

In a regime with $\Delta_l(H\to 0) \to 0$, we can substitute the derivative w.r.t.\ $H$ in \cref{eq:T-scaling} by $1/H$. From the integral representation of the (renormalised) chiral condensate, see e.g.\ \cite{Gao:2020fbl, Gao:2020qsj, Gao:2021wun} we can safely assume that $M_q \propto \Delta_l$, which also follows straightforwardly within dynamical hadronisation, see \cite{Fu:2019hdw}. We also use the relation between chiral condensate and pion mass, \cref{eq:mpi}. Note that the latter already implies in the limit $H\to 0$, that $\Delta_l \propto H^{r_1}$ with $r_1 < 1$. This follows from $m^2_\pi\propto H/\Delta_l \to 0$.

With the relations discussed above, \cref{eq:T-scaling} turns into,
\begin{align}\label{eq:TscalingExp}
	\Delta_l \propto H^\frac12 \,\frac{\Delta_l^3}{ \left(c\,H^{1/2}  + \Delta_l^{3/2}\right)^3}\,,
\end{align}
for $\Delta_l\to 0$. The constant $c$ in \cref{eq:TscalingExp} is irrelevant for the following discussion of the leading scaling behaviour.

We now proceed with a detailed analysis of the consequences of \cref{eq:T-scaling} for the $H$-scaling of the chiral condensate for small $H\to 0$: \\[-1ex]

\textit{$M_{q,\chi}>0$:} This condition is tantamount to large spontaneous symmetry breaking, and hence also a large chiral condensate. This situation holds true for temperatures smaller and sufficiently far from the chiral transition temperature. There, the constituent quark mass $M_q$ as well as the chiral condensate $\Delta_l$ have a sizeable non-vanishing part for $H\to 0$. Then, \cref{eq:T-scaling} reduces to
\begin{align}\label{eq:T-scalingTrivial}
	\frac{\partial \Delta_l}{\partial H}\propto \frac{1}{m_\pi}\,.
\end{align}
Moreover, the pion mass is proportional to $H^{1/2}$ as in the vacuum, see \cref{eq:vacChiral} and \cref{eq:mpi}. This leads us to
\begin{align}\label{eq:1/2}
\Delta_l\propto c_0 + c_{\frac12} H^\frac12\,,
\end{align}
that is $r_1=1/2$. This scaling is the analogue of the $H\log H$ scaling in the vacuum, the difference coming from the dimensional reduction 4d$\to$3d present in the Matsubara zero mode. Clearly, in this regime quasi-massless modes are present, as it solely originates in the small pion mass in comparison to other mass scales. We close the discussion of this regime with the remark, that in this limit we also expect a chiral scaling of the vertices in the diagrams for the chiral susceptibility with powers or inverse powers of $m_\pi^2$. This originates in the infrared singularities similar to those leading to the $1/m_\pi$ scaling in \cref{eq:T-scalingTrivial}. This may change the scaling \cref{eq:1/2}. Indeed, the latter naively relates to a $\Delta_l^3$ term in $V_\textrm{eff}$, which is at odds with the symmetry of the effective potential under $\Delta_l\to -\Delta_l$.     \\[-1ex]

\textit{$M_{q,\chi}\approx 0$:}  Sufficiently close to the chiral phase transition, both the constituent quark mass $M_q$ as well as the chiral condensate melt away. For $r_1\leq 1/3$, which potentially includes the mean field scaling, we have $c\,H^\frac12+ \Delta_l^2\propto \Delta_l^2$ in the limit $H\to 0$. Then, \cref{eq:TscalingExp} leads to
\begin{align}\label{eq:1/5}
		\Delta_l \propto H^\frac15\,.
\end{align}
\Cref{eq:1/5} comprises the consistent dimensional scaling for small spontaneous chiral symmetry breaking scales at finite temperature. In terms of the order parameter potential it relates to a monomial $\Delta_l^6$ potential, see \cref{eq:VnScaling}, and hence higher order scatterings of the scalar-pseudoscalar channel. Note that this scaling is also the critical one in approximations with vanishing anomalous dimension (of the pion), $\eta=0$, see \cref{eq:delta}. \\[-1ex]

In summary, the scalings \cref{eq:MF}, \cref{eq:1/2} and \cref{eq:1/5} indicate the presence of quasi-massless modes. Finally, in the critical region, $r_1=1/\delta$ and $r_2$ is the subleading scaling coefficient The critical regime has already been discussed in detail in \Cref{sec:SizeCrit} and we refer the reader to this Section. Naturally, quasi-massless modes are present in the critical regime.\\[-1ex]

This concludes our analysis of the potential scalings of the chiral condensate with the light current quark mass: In regimes with the scalings \cref{eq:MF}, \cref{eq:1/2}, \cref{eq:1/5} and the critical scaling \cref{eq:delta}  quasi-massless modes are present, admitting chiral expansion schemes. This supports the construction of phenomenological chiral models in this regime.

\subsection{Validity regime of chiral expansions}\label{sec:QuasiMasslessResults}

\begin{figure}[t]
	\includegraphics[width=.93\columnwidth]{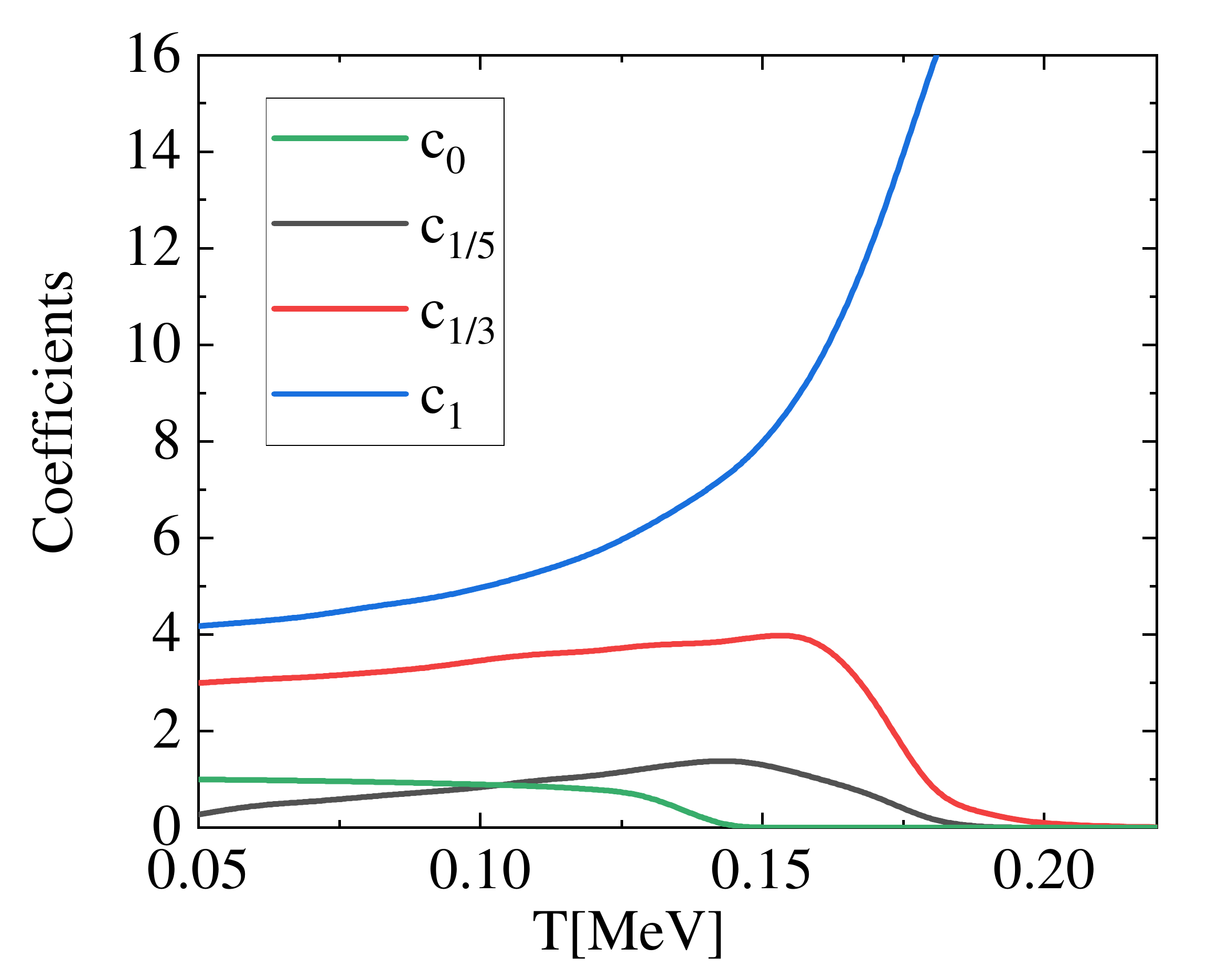}
		\caption{Temperature dependence of the coefficients $c_0,c_1,c_\frac13, c_\frac15$ in the global fit \cref{eq:FlobalFitDelta} for $\Delta_l(T,H)$.}\label{fig:cCoeff}
\end{figure}

In summary we are led to a global fit for the chiral condensate $\Delta_l(T,H)$ in the validity range of chiral expansions, excluding the $\chi$PT region. In this regime \Cref{eq:Ratxpanse} with $r_1=1/5$ and $r_2=1/3$ is valid, as already argued in \Cref{sec:Quasi-Massless}.
This reads,
\begin{align} \label{eq:FlobalFitDelta}
	\Delta_l(T,H) \approx &\,\Delta_{l,\chi}(0)\Bigl(c_0 +  c_\frac15 H^\frac15+ c_\frac13 H^\frac13+c_1 H\Bigr)\,,
\end{align}
and the temperature-dependent coefficients $c_i(T)$ with $i=0,1/3,1/5,1$ are depicted in \Cref{fig:cCoeff}. The different terms follow from a polynomial potential: the linear $H$ term from $\Delta_l^2$,  the $H^{1/3}$ term from $\Delta_l^4$, and the $H^{1/5}$ term from $\Delta_l^6$, in combination this leads us to the potential \cref{eq:Potbreaking} already discussed before.

The highest order term $\Delta_l^6$ in this potential relates to a stronger chiral dynamics including multi-scatterings. This terms dominates for small spontaneous symmetry breaking close to the chiral transition temperature $T_{c0}$, and can be inferred from the scaling  analysis of the susceptibility, see also the discussion about \Cref{eq:1/5}. In conclusion, as already discussed below \cref{eq:VnScaling}, we expect the higher order terms in the potential to be relevant for small $\Delta_l$ and hence small $H$. Therefore, global fits to $\Delta_l(T,H)$ may work well.

The above analytic analysis is confirmed within a numerical leading order analysis for smaller quark masses and close to the chiral transition temperature. The scaling coefficient $r_1$ as well as $\beta$ have been determined from our numerical data as
\begin{align}\label{eq:ScalingQuasiLess}
	\beta={ 0.400} ({ 9})\,,\qquad r_1={ 0.200} ({ 6})\,,
\end{align}
with $\beta$ being extracted from $c_0(T)$, as $\Delta_{l,\chi}(T)= \Delta_{l,\chi}(0) \,c_0$ for $T\leq T_{c0}$. In this regime the coefficient $c_1(T)$ of  the linear $H$ term decreases towards the chiral transition temperature $T_{c0}$, where it vanishes, see \Cref{fig:cCoeff}. We emphasise that the scaling \cref{eq:ScalingQuasiLess} in this regime with $r=0.2$ is close to, but distinct from the critical scaling, $r_{O(4)}\approx 0.207$ for $O(4)$-scaling and  $r_{O(2)}\approx 0.209$ for $O(2)$-scaling. This can be inferred from the respective $\chi^2$. We rush to add, that within the present approximation we do not expect full critical scaling in the critical regime, as discussed before. However, in the corresponding fRG analysis in \cite{Braun:2020ada} even a polynomial scaling with $r=1/5$ can be distinguished from the critical scaling with either $r_{O(4)}$ or $r_{O(2)}$. For the meson masses considered there ($m_\pi\gtrsim 30$\,MeV, that is $m_l\gtrsim 0.1$\,MeV), no critical scaling but polynomial scaling has been found.

The coefficients in \Cref{eq:FlobalFitDelta} are now obtained from $H$-fits in regimes $0\leq H\leq H_\textrm{max}$ for $\Delta_l(T,H)$ for a given temperature $T$. Here, $H_\textrm{max}$ is the maximal $H$ for which the coefficients as well as the $\chi^2$ error is stable. In turn, the complementary regime (for $H\leq 1$)  with $H_\textrm{max} < H \leq 1$ has a linear $H$ scaling and does not admit a chiral expansion.
\begin{figure}[t]
	\includegraphics[width=.98\columnwidth]{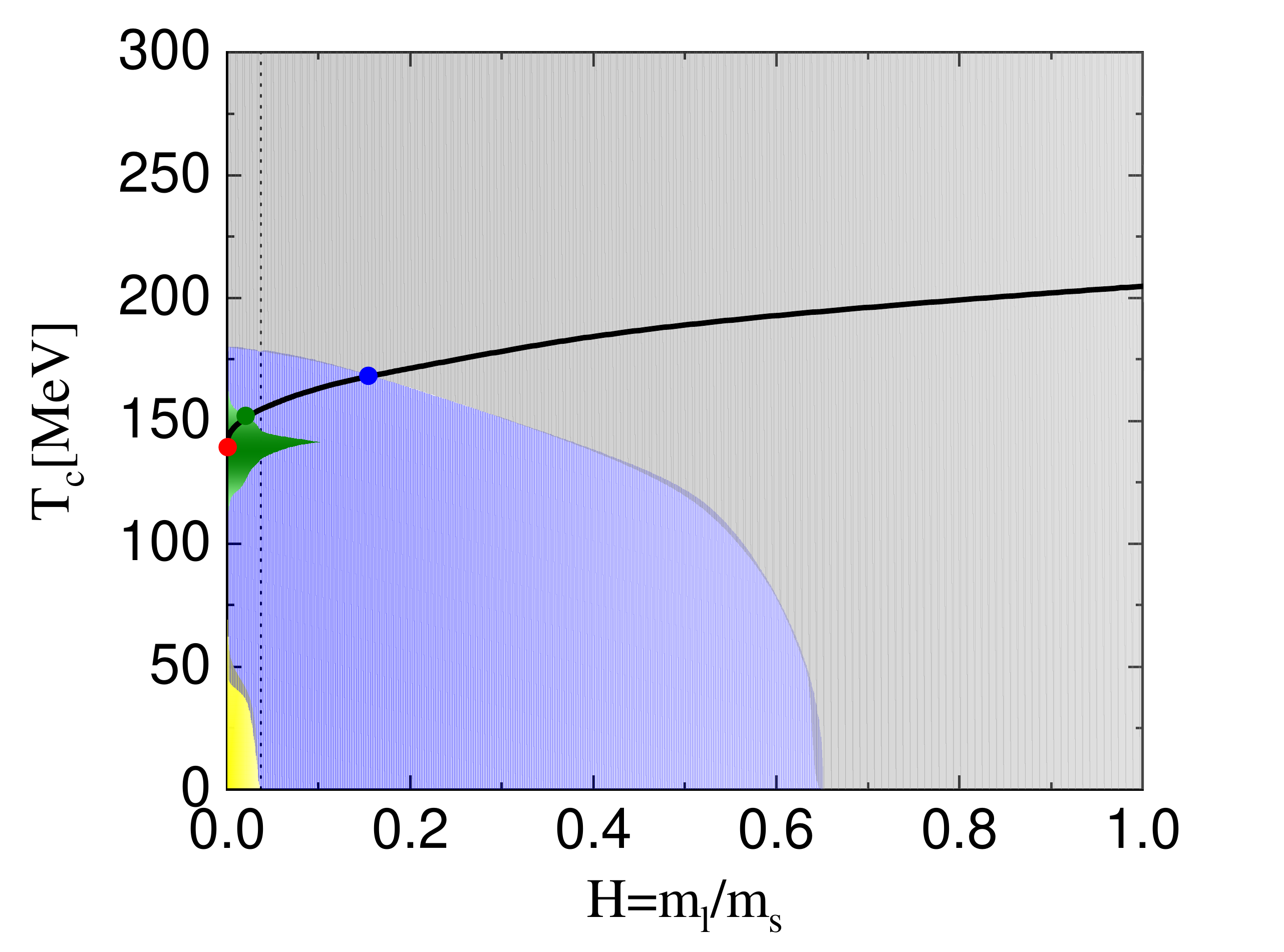}
	\caption{Validity range of chiral expansions in the current quark mass and temperature plane: \textit{yellow} (vacuum $\chi$PT ), \textit{blue} (dominating $H^{1/3}$-scaling), \textit{green} (dominating $H^{1/5}$-scaling). For the definition of dominating scaling see \Cref{sec:QuasiMasslessResults}. The \textit{grey} area indicates the linear $H$ scaling for large $m_l$ (no chiral expansion). The \textit{green bullet} indicates the intersection point $(T\,,\,H) \approx (152\, {\rm MeV}\,,\, 0.023)$ with $m_\pi \approx 110$\,MeV of the \textit{green} regime with the chiral transition line. The \textit{blue bullet} indicates the intersection point $(T\,,\,H) \approx (166\, {\rm MeV}\,,\, 0.14)$ with $m_\pi \approx 270$\,MeV of the \textit{blue} regime with the chiral transition line. We have also include an estimate of the small critical scaling regime (\textit{red area}, discussed above \cref{eq:expan}). The vertical line indicates QCD with physical current quark masses, that is $H=m_l/m_s = 1/27$. }\label{fig:ChiralRange}
\end{figure}
The fit \Cref{eq:FlobalFitDelta} with the temperature dependent coefficients depicted in \Cref{fig:cCoeff} allows us to provide a more detailed depiction of all the different regimes. For the present qualitative analysis we define sub-regimes in the validity regime of chiral expansions with the relative dominance of the respective $c_{r_i} H^{r_i}$ term over the other term: $c_{r_i} H^{r_i}\geq c_{r_j} H^{r_j}$ for all $j=1/3,1/5,1$: The respective regimes  are indicated as (\textit{green} ($H^{1/5}$), \textit{blue}) ($H^{1/3}$), \textit{yellow} (vacuum $\chi$PT) in \Cref{fig:ChiralRange}. The regime with large light current quark masses is indicated with \textit{grey}.

Specifically, we find a large regime with dominant $H^{1/3}$-scaling. As discussed before, this behaviour originates from an order parameter potential with a $\Delta_l^4$ interaction term. A subleading Gau\ss ian $\Delta_l^2$-term leads to an additional linear term in $H$. The absence of higher order terms with $\Delta_l^{2n}$ with $n\geq 3$ in the full effective potential $V_\textrm{eff}$ indicates a weak chiral dynamics. Hence, while such a scaling resembles (critical) mean field scaling, we rather interpret this as trivial chiral scaling, related to a weak chiral dynamics. Evidently, the latter works in favour of the convergence of the expansion.

Most importantly, in our opinion, the presence or absence of critical scaling in a small critical regime is not relevant for the phenomenological application of chiral transport models or chiral fluid dynamics, however, the presence of quasi-massless modes is: clearly, in the regime with $H^{1/5}$-scaling (\textit{green} in \Cref{fig:ChiralRange}) quasi-massless modes are present ($\lambda_2\approx 0$) and dominate the dynamics. A detailed analysis of the respective order parameter potential at $T_{c0}$ has already been provided in \Cref{sec:OrderPot}, see in particular \Cref{fig:VTchi} and the related discussion. In particular, in this regime the lowest order scattering of the scalar interaction channel is absent and the dynamics is only carried by the $\Delta^6$-term. This suggests to study the impact of this dynamics on QCD transport.

\section{Summary}\label{sec:Summary}

In this work we have studied the magnetic equation of $2$- and $2+1$-flavour QCD, mostly concentrating on the latter physical case. This has been done in a generalised functional first principles  approach to QCD, set-up in \cite{Gao:2020qsj, Gao:2020fbl, Gao:2021wun}. There it has been used for the phase structure of QCD as well as precision computations in the vacuum. Importantly, in this approach no phenomenological parameters has to be included.

We have computed the quark condensate and chiral susceptibility for general light current quark masses, while keeping the strange quark mass fixed in $2+1$-flavour QCD. The magnetic equation of state and the light current quark mass dependence of the chiral transition temperature is in quantitative agreement with that from the recent functional renormalisation group study \cite{Braun:2020ada}, see in particular \Cref{fig:sus}, \Cref{fig:Tc} and \Cref{fig:TcMpi}. Moreover, the critical temperature in the chiral limit is given by $T_{c0}=141$\,MeV. We also discussed a combined systematic error estimate from the present functional results and the present work, as well as lattice results from \cite{Ding:2019prx, Kotov:2021rah}, leading to a range $132\,\textrm{MeV} \lesssim T_{c0} \lesssim  141\,\textrm{MeV}$, see \cref{eq:CombinedEstimate}. We stress that this estimate could be much reduced in a combined study with functional and lattice approaches, exploiting the respective strengths.

So far, functional computations suggest a very small scaling window with critical scaling from low energy effective theories, \cite{Braun:2007td, Braun:2009ruy, Braun:2010vd, Klein:2017shl}, as well as functional QCD, \cite{Braun:2020ada}, see \Cref{sec:ChiralMassless}. The present approximation does not incorporate the full back coupling of the chiral dynamics, and hence cannot add to this intricate question. However, it is an interesting observation, that the magnetic equation of state from the present DSE computation without the inclusion of the potential critical dynamics agrees quantitatively with that in \cite{Braun:2020ada}, where the potential critical dynamics is taken care of for $m_\pi\gtrsim 30$\,MeV. This hints at the quantitative irrelevance of the critical dynamics for a large range of pion masses.

More importantly, we have argued that for phenomenological applications to heavy ion physics in the vicinity of the chiral phase transitions the presence or absence of a large critical regime is not important. Instead, it is the presence of quasi-massless modes which allows for the use of chiral transport models or chiral fluid dynamics, e.g.~\cite{Bluhm:2018qkf, Bluhm:2020rha, Grossi:2021gqi, Florio:2021jlx}. These transport models require QCD input, most prominently in terms of the full dispersion, see \cite{Bluhm:2018qkf} for first steps in this direction, and the full interaction. In this paper we have contributed to the computation of the latter in computing the full chiral order parameter potential for the first time within the DSE approach, see \Cref{sec:OrderPot}, and in particular \Cref{fig:EffPot} and \Cref{fig:lambdai}. This computation utilises the light current quark mass dependence of the chiral condensate, whose knowledge also allows us to estimated the size of the regime with quasi-massless modes in \Cref{sec:ChiralMassless}, see in particular \Cref{fig:ChiralRange} and the respective discussion. Importantly, we find a large regime with quasi-massless modes. In particular we can identify the validity regime of vacuum $\chi$PT (\textrm{red} regime in \Cref{fig:ChiralRange}) and a regime about the phase transition temperature in the chiral limit, $T_{c0}$  (\textrm{green} regime in \Cref{fig:ChiralRange}, the respective potential is shown in \Cref{fig:VTchi}) which allow for a direct expansion about the chiral limit. Finally, we can identify a large regime with a Ginsburg Landau type order parameter potential  (\textrm{blue} regime in \Cref{fig:ChiralRange}), which also should allow for chiral expansions. The asymptotic regime with large current quark masses in indicated with \textrm{grey} regime in \Cref{fig:ChiralRange}.

The present results provide highly welcome support for the current use of chiral transport models and chiral fluid dynamics in heavy ion collisions at small densities. Currently, we refine the present qualitative analysis, as well as extending it to the high density regime including the potential CEP. We hope to report on the respective results in the near future.
\\[-1ex]

\noindent{\bf Acknowledgements}\\[-0ex]
We thank J.~Braun, G.~Eichmann, C.~F.~Fischer, W.-j.~Fu, A.~Y. Kotov, M.~P.~Lombardo, J.~Papavassiliou, F~.Rennecke, B.-J.~Schaefer, C.~Schmidt, A.~Trunin and N.~Wink for discussions. We also thank the other members of the fQCD collaboration \cite{fQCD} for discussions and collaboration on related subjects. F.~Gao is supported by the Alexander von Humboldt foundation. This work is supported by EMMI and the BMBF grant 05P18VHFCA. It is part of and supported by the DFG Collaborative Research Centre SFB 1225 (ISOQUANT) and the DFG under Germany's Excellence Strategy EXC - 2181/1 - 390900948 (the Heidelberg Excellence Cluster STRUCTURES).

\vfill

\bibliography{ref-lib}
\end{document}